\documentclass[prc,aps,nofootinbib,%
tightenlines,superscriptaddress,floatfix,
preprintnumbers]{revtex4}
\usepackage{bbm}
\usepackage{graphicx}
\usepackage{amsmath}
\usepackage{amsfonts,amsbsy}
\usepackage{amssymb}

\newcommand{\xbj}{{x}}
\newcommand{\dsigmap}{{\frac{\ud \sigma^\textrm{p}_\textrm{dip}}{\ud^2 \bt}}}
\newcommand{\qscgc}{{\bar Q}_\mathrm{s}}
\newcommand{\ud}{\, \mathrm{d}}
\newcommand{\qssp}{{Q^2_\mathrm{s,p}}}

\newcommand{\as}{{\alpha_{\mathrm{s}}}}
\newcommand{\rt}{{\mathbf{r}_\perp}}
\newcommand{\ra}{{\mathbf r}_1}
\newcommand{\rb}{{\mathbf r}_2}

\newcommand{\bt}{{\mathbf{b}_\perp}}

\begin{document}
\title{Saturation models of HERA DIS data and inclusive hadron distributions\\ in p+p collisions at the LHC}
\author{Prithwish Tribedy}
\affiliation{Variable Energy Cyclotron Centre, 1/AF Bidhan Nagar, Kolkata-700064, India}
\author{Raju Venugopalan}
\affiliation{Physics Dept., Bldg. 510A, Brookhaven National Laboratory, Upton, NY 11973, USA}

\begin{abstract}
Unintegrated gluon distributions sensitive to the transverse spatial distribution of gluons in the proton are extracted from data on exclusive and diffractive final states at HERA in the dipole approach. These unintegrated gluon distributions can be used to compute inclusive hadron production in p+p collisions at the LHC. In this paper, we consider a number of saturation models with differing dynamical assumptions that give good fits to the available HERA data. We apply these models to study the rapidity and transverse momentum dependence of the LHC data up to $\sqrt{s} = 7$ TeV. We examine the 
sensitivity of these results to parameters that are not constrained by the HERA data and comment on similarities and differences with previous work. We compute the n-particle inclusive multiplicity distribution and show that the LHC p+p results are in agreement with predictions for multi-particle production in the Color Glass Condensate approach. This result has significant ramifications for the interpretation of multi-particle correlations in high multiplicity events at the LHC.

\end{abstract}

\preprint{ }

\maketitle

\section{Introduction}

Colliders at ultra-high energies have made feasible the study of the collective many-body properties of QCD in the limit of fixed $Q^2\gg \Lambda_{\rm QCD}^2$ and very small $x$. An important property of QCD in these ``Regge asymptotics" is 
the phenomenon of gluon saturation, which states that a probe with transverse resolution $1/Q^2$ interacts with a target with probability of order unity when $Q^2 \leq Q_S^2(x)$, where $Q_S^2(x)$ is a dynamically generated semi-hard scale in the target hadron or nucleus~\cite{GLR}. In the framework where the parton model is manifest, the gluon saturation phenomenon corresponds to the requirement that transverse momentum modes with $k_\perp \leq Q_S$ are maximally occupied with an occupation number $1/\alpha_S$. The large occupation number means that these modes can be treated classically~\cite{MV}; an additional separation of scale between fast and slow modes due to time dilation allows for a renormalization group treatment of these high occupation number states as they evolve with energy~\cite{JIMWLK}. This Color Glass Condensate (CGC)~\cite{CGC-reviews} description of saturated gluons is universal to hadrons and nuclei and can be tested both in deeply inelastic scattering experiments off hadrons and nuclei as well as in hadron/nuclear collisions. 

The CGC approach, when applied to Deeply Inelastic Scattering (DIS), results at leading order in 
$\as$~\cite{McLerran:1998nk} in the dipole picture where the inclusive virtual photon hadron cross section is expressed as~\cite{NikZak}
\begin{equation}\label{eq:sigmatot}
\sigma^{\gamma^*p}_{L,T}
= \int\! \ud^2 \rt \int_0^1 \! \ud z \left| \Psi^{\gamma^*}_{L,T}
\right|^2 
\int \! \ud^2 \bt \dsigmap    .
\end{equation}
Here $\left| \Psi_{L,T}^{\gamma^*}(\rt,z,Q) \right|^2$
represents the probability for a  virtual photon to produce a quark--anti-quark pair of size $r = |\rt|$ and $\dsigmap(\rt,\xbj,\bt)$ denotes the \emph{dipole cross section} for this pair to scatter off the target at an impact parameter $\bt$. The former is well known from QED, while the latter represents the dynamics of QCD scattering at small $x$. It was shown some time ago that a model known as the Golec-Biernat--Wusthoff  (GBW) model~\cite{GolecW} that implements saturation in the dipole cross-section through the parametrization $\dsigmap = 2 ( 1 - e^{ - r^2 \qssp(x)/4})$, where $\qssp (x) = (x_0/x)^\lambda$ GeV$^2$, gives a good qualitative fit to the HERA inclusive and diffractive cross section data for $x_0 = 3\cdot 10^{-4}$ and $\lambda = 0.288$. This work explained very simply key features of the HERA data and was very suggestive of the possible role of a semi-hard saturation scale in the hadron. 
This work was further refined in more sophisticated models that are somewhat better motivated and treat the impact parameter dependence of the dipole cross-section more accurately. As we shall discuss further in the next section, these models give excellent fits to small x inclusive, diffractive and exclusive HERA data. The common ingredient in these combined fits is the dipole cross-section.

The dipole cross-section, to leading logarithmic accuracy, is a universal quantity which can be applied to compute inclusive quantities in hadron-hadron collisions. It is defined in terms of the real part of the forward scattering amplitude ${\cal N}(\rt,x,\bt)$ as 
\begin{equation}
\dsigmap(\rt,x,\bt)
= 2 \,\, {\cal N}(\rt,x,\bt) \equiv
2\,\left(1-\frac{1}{N_c}\,\Big< {\rm tr}\, \big({\tilde U}(\bt + \frac{\rt}{2}){\tilde U}^\dagger(\bt-\frac{\rt}{2})\big)\Big>_x\;\right)\; ,
\label{eq:Wilson-amp}
\end{equation}
where ${\tilde U}(\bt \pm \frac{\rt}{2})$ is a Wilson line in the fundamental representation representing the interaction between a quark and the color fields of the target. The average $\langle\cdots\rangle_x$ is an average over these color fields; the energy dependence of the correlator as a function of x (or the rapidity $Y=\ln(1/x)$) is given by the JIMWLK equation~\cite{JIMWLK}. In the large $N_c$ limit, the equation for the energy evolution of this correlator is the Balitsky-Kovchegov (BK) equation~\cite{BK}. We note however that neither JIMWLK nor BK is at present equipped to deal well with the impact parameter dependence of the dipole 
cross-section; the dipole cross-section in this formalism is 
taken in eq.~(\ref{eq:Wilson-amp}) to be independent of the impact parameter. To address the impact parameter dependence of this equation, one resorts to models which parametrize both saturation effects and the impact parameter dependence.

In hadron-hadron collisions, one can derive at leading order the expression~\cite{BGV1}
\begin{equation}
 \frac{\textmd{d}N_{g}(\textbf{b}_{\bot})}{\textmd{d}y~\textmd{d}^{2}\textbf{p}_{\bot}}=\frac{16 \alpha_S}{\pi C_F} \frac{1}{p_{\bot}^2} 
\int \frac{\textmd{d}^{2} \textbf{k}_{\bot}}{(2\pi)^{5}} \int \textmd{d}^{2} \textbf{s}_{\bot} \frac{\textmd{d}\phi_A(x_1,\textbf{k}_{\bot}|\textbf{s}_{\bot})}{\textmd{d}^2\textbf{s}_{\bot}}
\frac{\textmd{d}\phi_B(x_2,\textbf{p}_{\bot}-\textbf{k}_{\bot}|\textbf{s}_{\bot}-\textbf{b}_{\bot})}{\textmd{d}^2\textbf{s}_{\bot}}
\label{eq:ktfact1}
\end{equation}
This equation is a generalization of the well known $k_\perp$ factorization expression for inclusive gluon production~\cite{Braun} to include the impact parameter dependence of the unintegrated gluon distributions. Here $C_F = N_c^2-1/2 N_c$ is the Casimir for the fundamental representation. Using a relation between quark and gluon dipole amplitudes strictly valid in the  large $N_c$ limit, the unintegrated gluon distribution in either of the two protons can be expressed in terms of the corresponding dipole cross-section measured in DIS as~\cite{GelisSV} 
\begin{equation}
\frac{\textmd{d}\phi(x,\textbf{k}_{\bot}|\textbf{s}_{\bot})}{\textmd{d}^2\textbf{s}_{\bot}} =\frac{\textbf{k}_\bot^2 N_c}{4 \alpha_S}  \int \limits_{0}^{+\infty}\textmd{d}^2\textbf{r}_{\bot}
e^{i \textbf k_{\bot}.\textbf{r}_{\bot}} \left[1 - \frac{1}{2}\, \frac{\ud \sigma^\textrm{p}_\textrm{dip}}{\ud^2 \textbf{s}_\perp\\} (\rt,x,\textbf{s}_\perp)\right]^{2}
\label{eq:unint-gluon}
\end{equation}
Thus the impact parameter dependent dipole cross-section determined from HERA data can be used to compute the single inclusive gluon distribution in proton-proton collisions with no additional parameters. This statement is strictly valid to leading log accuracy for momenta $k_\perp > \qssp$. Further, as we shall discuss later, there will be additional parameters that come in when one wants to make contact with the measured hadron spectrum. 

\begin{figure}[t]
\centerline{
\includegraphics[height=6cm,width =5cm]{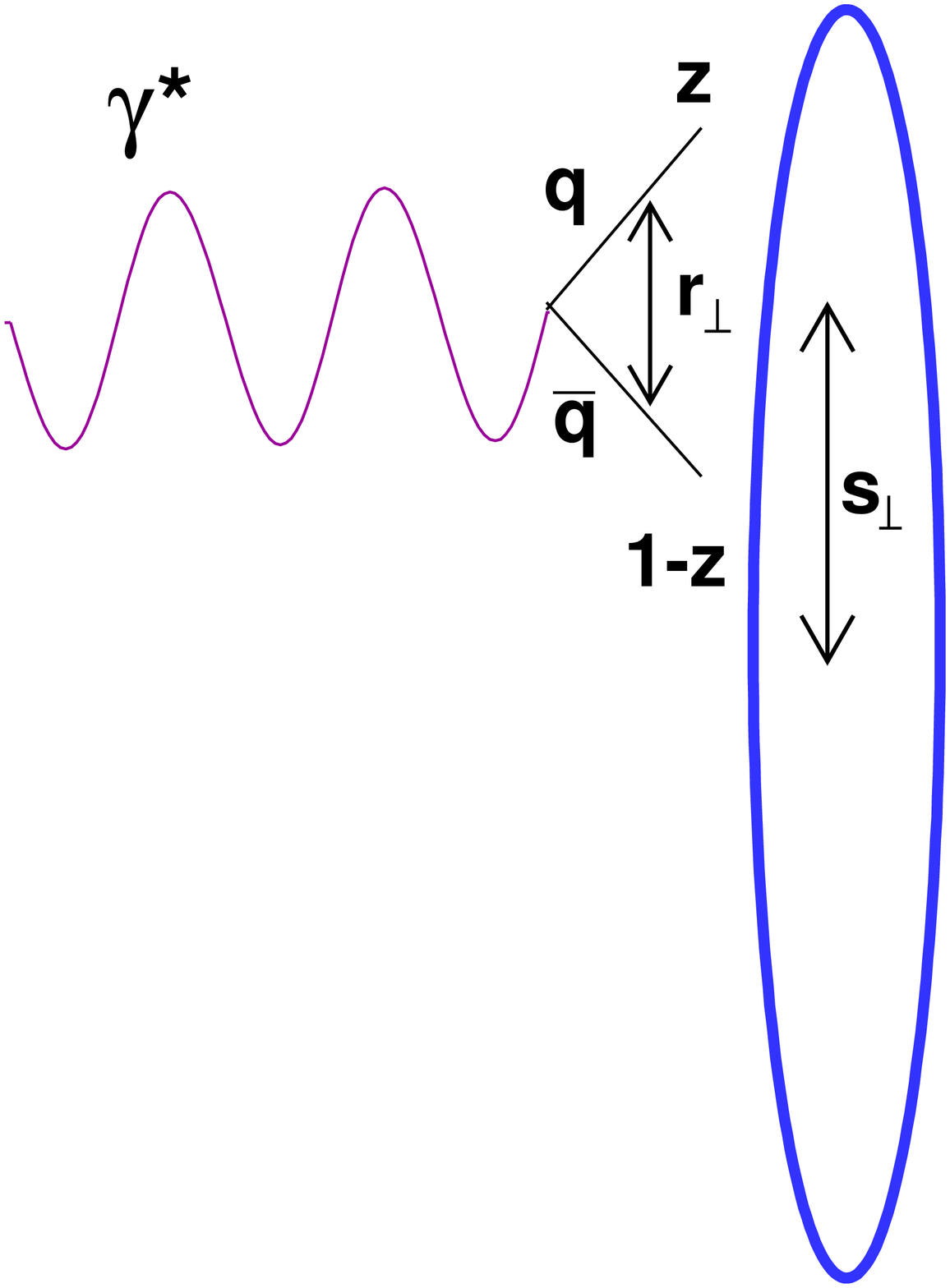}
\includegraphics[width =2cm]{vspace.eps}
\includegraphics[height=6cm,width =6cm]{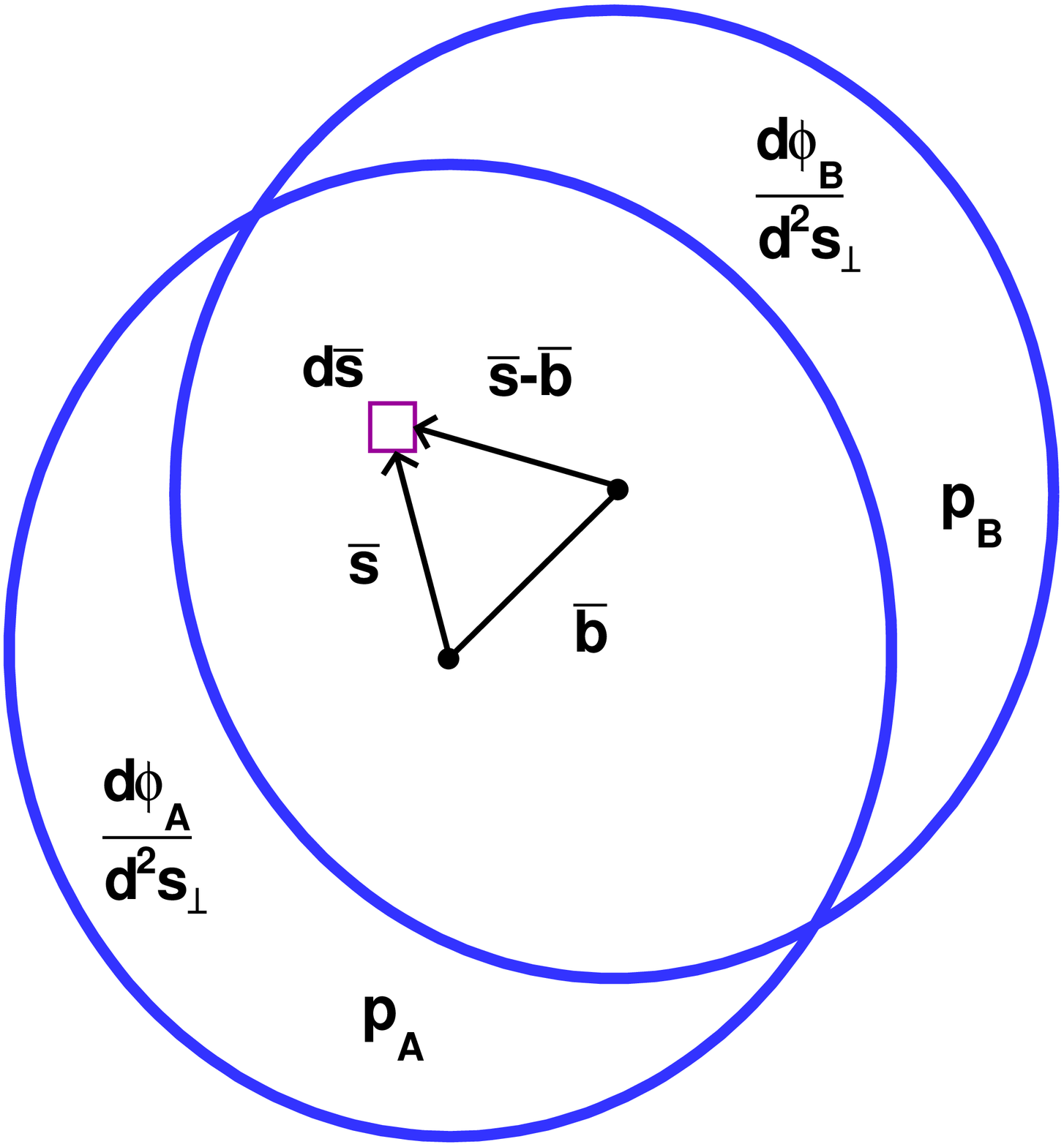}
}
\caption{Left: Dipole cross-section in DIS. Right: Overlap of unintegrated gluon distributions in proton-proton collisions.}
\label{fig:ppcol}
\end{figure}

This approach was applied most recently to compute the single inclusive hadron spectrum in proton-proton collisions at the LHC by Levin and Rezaeian~\cite{Lev-Rez}. The quantitative differences of our study to their work are the following:
a) we consider  three dipole models that give good fits to HERA data to see whether they give results consistent with the LHC data, b) we study and comment on the dependence of the results on variations of the parameters in the study and c) we convolve the inclusive gluon distribution with fragmentation function instead of using a simple fragmentation presciption as in ref.~\cite{Lev-Rez}. We shall also comment on other quantitative differences in our respective treatments at the appropriate points in the text. A qualitative difference of our work relative to that of ref.~\cite{Lev-Rez} is that we compute directly the average inclusive  multiplicity at a given impact parameter. In computing the minimum bias single inclusive multiplicity distribution, there are similar uncertainties as ref.~\cite{Lev-Rez}, which can be fixed by normalizing the data to single inclusive data at lower center of mass energies. However, as we shall discuss later, the average multiplicity at a given impact parameter is an essential input in computing the probability distribution as a function of event multiplicity. We shall compute the n-particle probability distribution and compare our results with the p+p collider data. These results will be important in understanding the role of various sources of fluctuations in the p+p collider data. 

\section{Proton dipole cross-section}

In this section, we shall discuss some of the dipole model parametrizations that have been compared extensively to the HERA data. We shall consider respectively the IP-Sat model~\cite{KT}, the b-CGC model~\cite{IIM,KMW,KW} and the NLO-BK model~\cite{Albacete}. Our list of dipole parametrizations is by no means exhaustive~\cite{dipole-HERA} but is a sample of some of the key approaches where extensive comparisons have been made to the data. We note that dipole models have been applied to understand RHIC data~\cite{KharzeevL}, and there have also been preliminary attempts at a combined analysis of RHIC, HERA and fixed target e+A data~\cite{Navarra1,AlbaceteM1,AlbaceteM2,KLV,DuslingGLV}. We also note that there are leading twist models with some similarities to dipole models that have been compared to the HERA data~\cite{FrankfurtGS}. 

\subsection{The IP-Sat Model}
The impact parameter dependent dipole saturation model (IP-Sat)~\cite{KT} is a refinement of the Golec-Biernat--Wusthoff dipole model~\cite{GolecW} to give the right perturbative limit when $\rt\rightarrow0$~\cite{BGBK}. It is equivalent to the expression derived in the classical effective theory of the CGC, to leading logarithmic accuracy~\cite{McLerran:1998nk}. The proton dipole cross-section in this model is expressed as 
\begin{eqnarray}
\dsigmap(\rt,x,\bt)&=&2\left[1-\exp\left(-\frac{\pi^{2}}{2N_{c}}\rt^{2}\alpha_{S}(\mu^{2}) xg(x,\mu^{2})T_p(\bt)\right)\right]\, .
\label{eq:ipsat-dipole}
\end{eqnarray}
Here the scale $\mu^{2}$ is related to dipole radius $\rt$ (see fig.~\ref{fig:ppcol}) as
\begin{equation}
\mu^{2}=\frac{4}{\rt^{2}}+\mu_{0}^{2}\,,
\end{equation}
where the leading order expression for the running coupling is 
\begin{equation}
 \alpha_{S}(\mu^{2})=\frac{12\pi}{(33-2n_{f})\log(\mu^{2}/\Lambda_{QCD}^{2})}
\end{equation} 
with $n_{f}$=3, $\Lambda_{QCD}$=0.2 GeV.
The model includes saturation as eikonalized power corrections to the DGLAP leading twist expression and may be valid in the regime where logs in $Q^2$ dominate logs in $x$. The saturation scale for a fixed impact parameter is determined self--consistently by requiring that the dipole amplitude (within brackets in eq.~\ref{eq:ipsat-dipole})  have the magnitude ${\cal N}(x,r_S,\bt)= 1-e^{-1/2}$, with $\qssp = 2/r_S^2$. We note that there is an overall logarithmic uncertainty in the determination of $\qssp (x,\bt)$.

For each value of the  dipole radius, the gluon density $xg(x,\mu^{2})$ is evolved from $\mu_{0}^{2}$ to $\mu^{2}$ using LO DGLAP evolution equation without quarks, 
\begin{equation}
\frac{\partial xg(x,\mu^{2})}{\partial \log \mu^{2}} = \frac{\alpha_{S}(\mu^{2})}{2\pi}\int \limits_{x}^{1}
dz P_{gg}(z)\frac{x}{z}g\left(\frac{x}{z},\mu^{2}\right) 
\end{equation}
Here the gluon splitting function with $n_{f}$ flavor and $C_{A}$=3 $\&$ $T_{R}$=1 is 
\begin{equation}
 P_{gg}(z)=6\left[\frac{z}{(1-z)+}+\frac{1-z}{z}+z(1-z)\right]+
\left(\frac{11}{2}-\frac{n_{f}}{3}\right)\delta(1-z)
\end{equation}
The initial gluon density at the scale $\mu^{2}_{0}$ is taken to be of the form
\begin{equation}
 xg(x,\mu^{2}_{0})=A_{g}x^{-\lambda_{g}}(1-x)^{5.6}
\end{equation}
An important feature of the IP-Sat model is the b-dependence of the dipole cross-section, which is introduced through a gluon density profile function $T(b)$. 
This profile function is normalized to unity and is chosen to have the Gaussian form  
\begin{equation}
 T_{p}(\bt)=\frac{1}{2\pi B_{G}} \exp\left({-\bt^{2}\over 2B_{G}}\right)\,,
\label{eq:IPsat-imp-par}
\end{equation}
where $B_G$ is a parameter fit to the HERA diffractive data. This corresponds to $\langle b^2\rangle = 2 B_G$, the average squared {\it gluonic} radius of the proton.

Sets of parameters obtained from optimal fits of the IP-Sat model to HERA data~\cite{KMW} are listed in table \ref{tab_ipsat}.
\begin{table}[ht]
\centering
\begin{tabular}{lllll}
\hline
\hline
$m_c$ & $B_G (\rm{GeV}^{-2})$ & $\mu_0 (\rm {GeV}^2)$& \,\,\,\,$A_g$ & \,\,\,\,\,$\lambda_g$\\
\hline
1.4 & \,\,\,\,4.0 & 1.17 & 2.55 & \,\,\,\,\,0.020\\
1.35 & \,\,\,\,4.0 & 1.20 & 2.51 & \,\,\,\,\,0.024\\
1.5 & \,\,\,\,4.0 & 0.77 & 2.64 & \,\,\,\,0.011\\
1.4 &  \,\,\,\,4.0 &1.50 & 3.61 & \,\,\,\,\,-0.118\\
\hline
\hline
\end{tabular}
\caption{Parameters of the IP-Sat model obtained from the fit to HERA data~\cite{KMW}. }
\label{tab_ipsat}
\end{table}
All data sets except the last use $m_{\rm u,d,s}=0.14$ GeV; the last set corresponds to $m_{\rm u,d,s}=0.05$ GeV. The parameters of the initial gluon distribution are determined from fits to 
the HERA $F_2$ data~\cite{ZEUS,H1} with a $\chi^2\sim 1$. For charm quarks, $x=x_{\rm bj}(1+4m_c^2/Q^2)$. The value of $B_G$ is determined primarily from the $t$-distributions of $J/\psi$ mesons measured by 
ZEUS~\cite{Zeus-Jpsi} and H1~\cite{H1-Jpsi}. With these parameters, excellent agreement is obtained with the HERA exclusive vector meson and DVCS data. For a detailed comparison of this model to the HERA data, we 
refer the reader to Ref.~\cite{KMW}.

\subsection{The b-CGC Model}

The IP-sat dipole model is applicable when leading logarithms in $Q^2$ dominate over leading logarithms in $x$. 
At very small $x$, quantum evolution in the CGC describing  both the bremsstrahlung limit of linear small $x$ evolution as well as nonlinear RG
evolution at high parton densities, combined with a realistic $b$-dependence, may be better captured in the bCGC model~\cite{IIM,KMW,KW}.  The proton dipole cross-section in this case is expressed as 
\begin{eqnarray}
\dsigmap(\rt,x,\bt)=2\times\left\{ \begin{array}{cccc}
{\cal N}_0 \left(\frac{\rt \qscgc }{2} \right)^{2\left(\gamma_s+\frac{1}{\kappa \lambda Y} \ln\left(\frac{2}{\rt \qscgc }\right)\right)} ~ & : &  \rt \qscgc \leq 2; \\
1-\exp\left(-A\, \ln^2 (B\,\rt \qscgc)\right) ~ & : &  \rt \qscgc > 2; \\
\end{array} \right.
\label{eq:bCGC}
\end{eqnarray}
In this model, in contrast to the IP-sat model, the impact parameter dependence is introduced though the quantity $\qscgc (x,\bt)$, defined as 
\begin{equation}
\qscgc(x,\bt)=\left(\frac{x_0}{x}\right)^{\lambda/2}\, \left[\exp\left(-\frac{\bt^2 }{2B_{\rm CGC}}\right)\right]^{\frac{1}{2\, \gamma_S}}
\end{equation}
As previously, for comparison of scales among different saturation models, the relevant saturation scale for a fixed impact parameter is determined self--consistently by requiring that the dipole amplitude (the expression to the right of the curly bracket in eq.~\ref{eq:bCGC})  have the magnitude ${\cal N}(x,r_S,\bt)= 1-e^{-1/2}$, with the saturation scale defined as $\qssp = 2/r_S^2$. The coefficients $A$ and $B$ are obtained by requiring the two asymptotic forms of the dipole cross-section and their first derivatives are continuous at $\rt\qscgc =2$:
\begin{equation}
 A=\frac{{\cal N}_0^2\gamma_S^2}{(1-{\cal N}_0)^2\ln(1-{\cal N}_0)}~~\&~~B=\frac{1}{2}(1-{\cal N}_0)^{-\frac{(1-{\cal N}_0)}{{\cal N}_0\gamma_S}}
\end{equation}
The parameter $\kappa=9.9$ is fixed from the leading order BFKL value for this quantity

Table~\ref{tab_bcgc} presents the parameters of the model that are fitted to the HERA data~\cite{KW}. The parameter $B_{\rm CGC}$ is determined from the $t$-dependence of exclusive $J/\psi$ photo-production. For the b-CGC model, this parameter however cannot be easily interpreted as giving the square mean gluonic radius of the proton. The parameters presented in the table in the first and fourth lines do not give good fits to the data. The fit corresponding to the second line of the table gives the best fit to the data with $\chi^2\sim 1$. The third line of the table corresponds to a fit where no saturation form is employed (namely, the perturbative expression, without the diffusion term proportional to $Y=\ln(x_0/x)$, is extended to $\rt\qscgc \geq 2$); it gives equally good fits to the data. However, it should be noted that this choice of parameters will violate perturbative unitarity for large dipole sizes $\rt\>> 1/\qscgc$.

\begin{figure}[h]
\centerline{
\includegraphics[height=7cm,width =7cm]{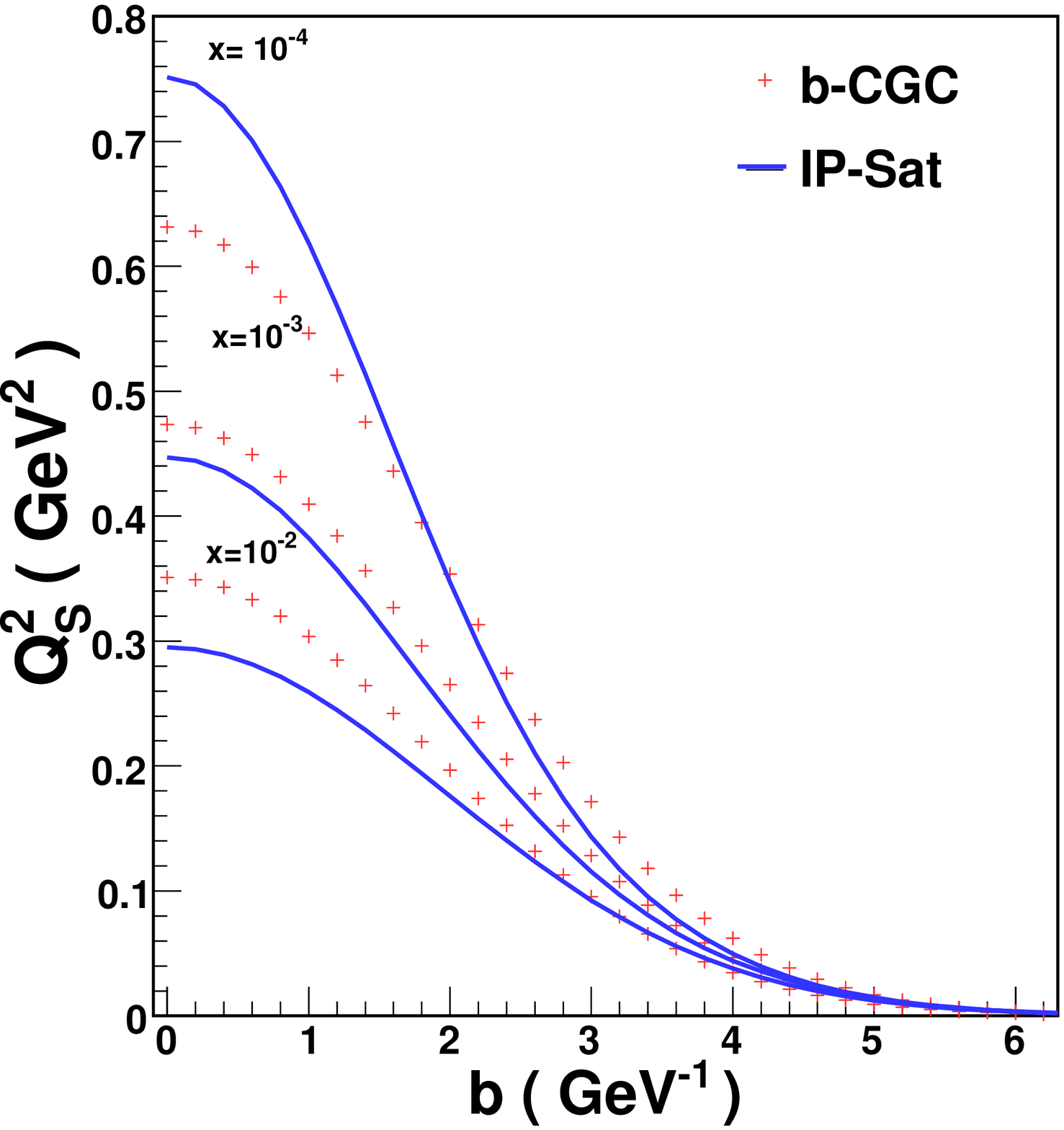}
\includegraphics[height=7cm,width =7cm]{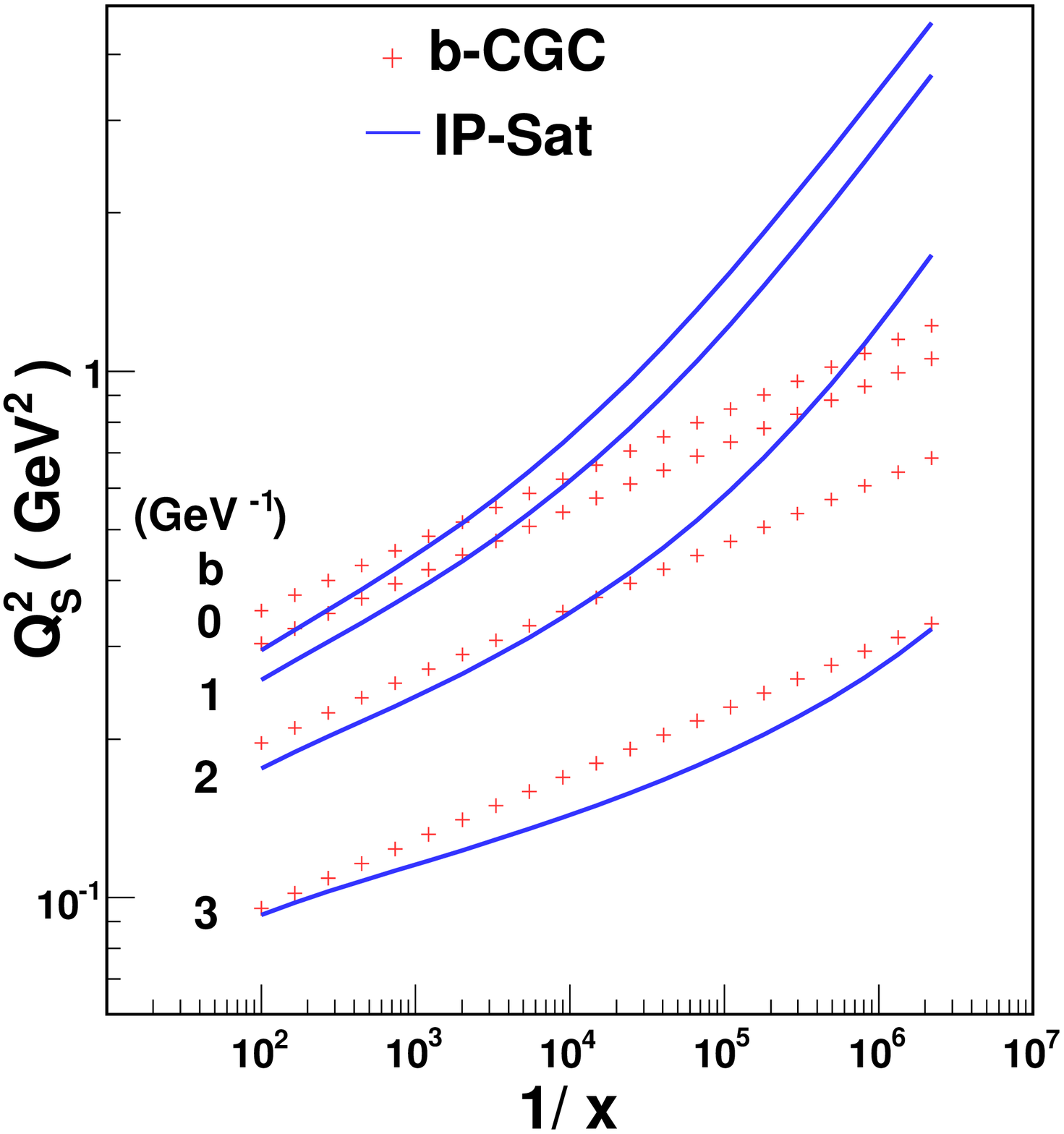}
}
\caption{ Saturation scales obtained from the IP-Sat(blue lines) and b-CGC models(red crosses) using first set of parameters in table.~\ref{tab_ipsat} and the second set in table.~\ref{tab_bcgc}. Left: impact parameter dependence of saturation scale different values of x. Right: x dependence of saturation scale for different values of impact parameter b.}
\label{fig:satscale}
\end{figure}

 The dependence of the saturation scale $\qssp(x,\bt)$ as a function of $x$ for different $\bt$ (and vice versa) in the IP-Sat and b-CGC models is shown in fig.~\ref{fig:satscale}. In both cases, the fits to the HERA data result in  a semi-hard scale ($Q_S^2 \gg \Lambda_{\rm QCD}^2$) with decreasing $x$ and $b$ values probed in the collisions. The existence of such scales and their increase with energy is what validates the whole approach of treating high parton densities in weak coupling. It would of course be naive to interpret the extracted numerical value of $Q_S$ as being precisely the scale that controls the running of the coupling. As is well known, the scale that controls the running of the coupling can differ considerably from this ``bare" scale in a given scheme for any given process. 

\begin{table}[ht]
\centering
\begin{tabular}{lllll}
\hline
\hline
\,\,\,
$\gamma_s$& $B_{\rm CGC} ({\rm GeV}^{-2})$ & ${\cal N}_0$&\,\, $x_0$ &\,\, $\lambda$\\
\hline
0.63 &\,\,\,\, 5.5 & 0.417 &\,\, 5.95$\cdot10^{-4}$ & 0.159\\
0.46 &\,\,\,\, 7.5 & 0.558 & \,\,1.84$\cdot10^{-6}$ & 0.119\\
0.43 &\,\,\,\, 7.5 & 0.565 & \,\,1.34$\cdot10^{-6}$ & 0.109\\
0.54 & \,\,\,\, 6.5 & 0.484 & \,\,3.42$\cdot10^{-5}$ & 0.149\\
\hline
\hline
\end{tabular}
\caption{Parameters of the b-CGC model obtained from fits to HERA data~\cite{KW}.The second row of parameters gives the best fit to HERA data.}
\label{tab_bcgc}
\end{table}


\subsection{The NLO-BK Model}

The dipole cross-section is defined in eq.~\ref{eq:Wilson-amp} as twice the forward scattering amplitude ${\cal N}$, which in the NLO-BK model~\cite{Albacete} satisfies the equation, 
\setlength\arraycolsep{.1pt}
\begin{eqnarray}
\frac{\partial {\cal N} (\rt,Y)}{\partial Y} 
= 
 \int {\ud\ra}\; {\mathcal K}_{_{\rm NLO}}(\rt,\ra,\rb)\times\big[ {\cal N}(\ra,Y) + {\cal N}(\rb,Y) - {\cal N}(\rt,Y) - {\cal N}(\ra,Y)\,{\cal N}(\rb,Y)\big] \; ,
\label{eq:BK-NLO}
\end{eqnarray}
\setlength\arraycolsep{1.4pt}
where $\rb \equiv \rt- \ra$ and the NLO-kernel (where running coupling corrections are taken into account) is given by
\begin{equation}
{\mathcal K}_{\rm NLO}(\rt,\ra,\rb) = \frac{\alpha_S(\rt) N_c}{\pi}\left[ \frac{\rt^2}{\ra^2 \rb^2} +\frac{1}{\ra^2}\left(\frac{\alpha_S(\ra^2)}{\alpha_S(\rb^2)}-1\right)+\frac{1}{\rb^2}\left(\frac{\alpha_S(\rb^2)}{\alpha_S(\ra^2)}-1\right)\right] \, .
\label{eq:NLO-BFKL-kernel}
\end{equation}
This expression is based on considerable recent work to include running coupling corrections to the BK equation~\cite{Balit2007,KovchW}. It should be noted that the expression does not include other next-to-leading log contributions to the 
kernel that have been computed recently~\cite{BalitC}. Also of relevance to us is the assumption in the evolution equation in eq.~\ref{eq:BK-NLO} that the dependence on the impact parameter and the dipole size factorize in the 
dipole amplitude as 
\begin{eqnarray}
{\cal N} (\rt,x,\bt)=2 \, {\cal T} (\bt)\,{\cal N} (\rt,x)\, .
\end{eqnarray}
The factorization here of impact parameter dependence and dipole size is very problematic conceptually and is an important limitation in applying these approaches to comparisons with data, except perhaps for final states that have limited sensitivity to the impact parameter dependence. How to include the impact pararmeter dependence in the BK/JIMWLK equations is an open question of great interest--we refer the reader to ref.~\cite{Berger-Stasto} and references therein. 

The NLO-BK model was applied in refs.~\cite{AlbacAMS1} to a phenomenological study of the HERA data on the proton structure function $F_2$.  Here the impact parameter dependence is taken to be a step function.  Two different models for the initial condition were used in that work. The first is the GBW model
\begin{eqnarray}
N(r,Y=0)=1-\exp\left[-\frac{Q^2_{s_0}r^2}{4}\right]
\end{eqnarray}
and the other is the MV model~\cite{MV}
\begin{eqnarray}
N(r,Y=0)=1-\exp\left[-\left(\frac{Q^2_{s_0}r^2}{4}\right)^{\gamma}\ln\left(\frac{1}{r \Lambda_{QCD}}+e\right)\right]
\end{eqnarray}
The initial condition for protons was determined from a global fit of
$F_2$ data in the work of \cite{AlbacAMS1}. The fit parameters are summarized in the table~\ref{tab:albacete}.
\begin{table}[h]
\begin{center}
\begin{tabular}{l | c | c | c | c}
I.C. & $\sigma_p$ (fm$^2$) & $Q_{s0,p}^2$ (GeV$^2$) & $C^2$ & $\gamma$ \\
\hline
GBW & 3.159 & 0.24 & 5.3 & NA \\
MV & 3.277  & 0.15 & 6.5 & 1.13 \\
\end{tabular}
\end{center}
\caption{Parameters for the initial condition of the
 proton dipole cross section obtained in \protect\cite{AlbacAMS1}.}
\label{tab:albacete}
\end{table}
The parameter $C^2$ here is a scale that controls the running of the coupling in ref.~\cite{AlbacAMS1}. 
This parametrization was extended to nuclei and fit to fixed target e+A DIS data~\cite{DuslingGLV}. It was then applied to study long range correlations in A+A collisions~\cite{DuslingGLV}, single inclusive~\cite{AlbaceteM1} and 
double inclusive~\cite{AlbaceteM2} distributions in deuteron--gold collisions at RHIC. In this work, only the MV parametrization will be considered.

\section{Single inclusive distribution in p+p collisions}

We shall now discuss the computation of the single inclusive hadron distribution in proton-proton collisions at the LHC obtained from the different dipole parametrizations discussed in the previous section. These parametrizations,  when inserted in eqs.~\ref{eq:unint-gluon} and \ref{eq:ktfact1} respectively, give the single inclusive gluon distribution. While it might appear that all the parameters are fixed by the fits to the HERA data, a comparison with the single inclusive 
hadron distribution quickly makes clear that one requires further assumptions and parameters. In making comparisons to the collider inclusive hadron data, we shall discuss the sensitivity of our results to variations in these parameters. The additional assumptions and parameters are as follows:

\begin{itemize}

\item The dipole cross-sections are fit to HERA data for $x\leq 0.01$. One therefore needs to make an assumption for $\phi(x,k_{\bot},b_{\bot})$ for larger $x > x_0 = 0.01$ values that kinematic regions of the proton-proton data are sensitive to. 
We use the parametrization~\cite{GelisSV}
\begin{equation}
\phi(x,k_{\bot},b_{\bot})=\left(\frac{1-x}{1-x_{0}}\right)^{\beta}\left(\frac{x_{0}}{x}\right)^{\lambda_{0}}\phi(x_{0},k_{\bot},b_{\bot}),~ x>x_{0}.
\label{eq:largex}
\end{equation}
This parametrization of the large $x$ unintegrated gluon distribution is motivated by quark counting rules~\cite{counting-rules} with fixed $\beta=4$ and the parameter $\lambda_{0}$ which ranges from $0$--$0.2$ in fits to the data. 

\item 
The single inclusive gluon distribution in eq.~\ref{eq:ktfact1} has a logarithmic infrared divergence which can be regulated either by putting a cutoff  on lower limit of $p_\perp$ or replacing $p_{\perp }$ by $m_\perp=\sqrt{p_\perp^2+m^2}$ where the mass is a free parameter. The single inclusive $p_\perp$ distribution is sensitive to the choice of $m$, which is fixed to be the same for data at all energy ranges.

\item Eq.~\ref{eq:ktfact1} corresponds to the rapidity distribution of inclusive gluons at a fixed impact parameter. However, what is measured is the pseudo-rapidity distribution; the rapidity can be expressed in terms of the 
pseudo-rapidity most generally as 
\begin{equation}
y(\eta,p_{\bot},m)=\frac{1}{2}\left[\frac{\sqrt{m^2+p_{\bot}^2ch^2\eta}+p_{\bot}sh\eta}{\sqrt{m^2+p_{\bot}^2ch^2\eta}-p_{\bot}sh\eta}\right] \, .
\label{eq:Jacobian}
\end{equation}
The single inclusive distribution with respect to the pseudo-rapidity therefore contains a Jacobian from the conversion of the expression with respect to the rapidity. We choose for economy of parameters the mass term in eq.~\ref{eq:Jacobian} to be the same as the one that regulates the infrared divergence. 

\item The minimum-bias single inclusive gluon distribution is obtained from the expression
\begin{equation}
\frac{d{\bar N}_{g}}{d^2p_{\bot}dy}=\frac{\int d^2\bt\frac{dN_{g}}{d^2p_{\bot}dy}(\bt) }{\int d^2\bt}
\label{eq:min-bias}
\end{equation}
 In the IP-Sat model, the proton profile does not change with energy. The transverse diffusion of the proton, often termed Gribov diffusion~\cite{Gribov}, is not fully accounted for by the diffusion of the unintegrated single inclusive gluon distribution because 
this growth does not automatically ensure the proper growth of the inelastic cross-section. Towards this end, we parametrize the maximum limit of b-integration to take the form $b_{\rm max.} = b_0 + C\,\ln(s)$. Fitting the available data on average dN/d$\eta$ as a function of energy we can extract $b_0$ and $C$. For first set of parameters in table~\ref{tab_ipsat}, the IP-Sat model gives $b_0$=5.17 GeV$^{-1}$ and C=0.19 with a choice of the mass term=0.4 GeV used in the Jacobian. This value of $b_0$ is close to $2 \,b_{\rm rms.}$ in the IP-Sat model, where $b_{\rm rms.}$ is the root mean square gluonic radius of the proton. The quantity $\pi b_{max}^2$, which is the denominator of eq.~\ref{eq:min-bias} can be interpreted as being closely related to the inelastic cross section contributing to particle production. Fig.~\ref{fig_sigmain} shows the variation of $\pi b_{max}^2$ in the IP-Sat model as a function of collision energy. These numbers are in the ballpark of estimates of the inelastic cross-sections at the LHC~\cite{D'Enterria}--they are however significantly higher than the values at lower energies. This is because the numbers for $b_{\rm max.}$ are extracted from fits  of eq.~\ref{eq:min-bias} 
to data; it is therefore also sensitive to the uncertainties in the numerator of eq.~\ref{eq:min-bias}. A possible interpretation is that these uncertainties are larger at lower energies thereby leading to an overestimate of the inelastic cross-section. 

An additional point with regard to fig.~\ref{fig_sigmain} is that one observes a $\sim 15\%$ difference of $\pi b_{\rm max.}^2$ at the highest LHC energies and may thereby hope to constrain the parameter $m$. However, due to the uncertainty in distinguishing the genuine inelastic cross-section from the NSD cross-section, it is unlikely at present that the 15\% difference can be definitive in that regard. A similar form of $b_{max}$ when used to fit average dN/d$\eta$ for b-CGC model gives C $\sim 0$. This suggests that because the impact parameter dependence of the dipole cross-section in the b-CGC model is tied in with its $x$ dependence (see eq.~\ref{eq:bCGC}), the non-trivial relation of the two in this model may well approximate the physics of Gribov diffusion. 

\begin{figure}[h]
\centerline{
\includegraphics[width =7cm, height =7cm]{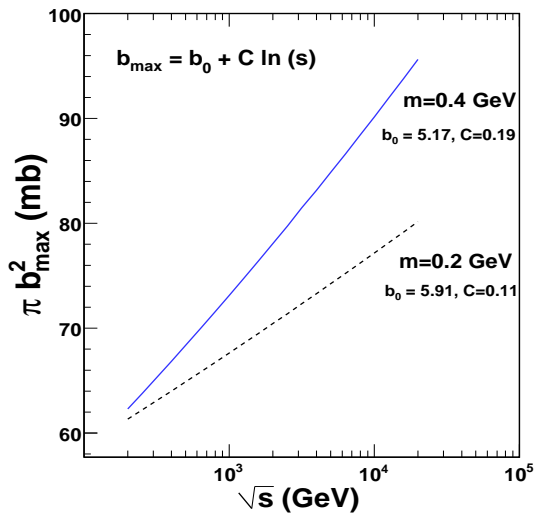}
}
\caption{Variation of the inelastic cross section with collision energy as obtained in the IP-Sat model for two different values of the infrared cut-off. See text for a detailed discussion.}
\label{fig_sigmain}
\end{figure}

\item The single inclusive hadron distribution is obtained by convolving eq.~\ref{eq:min-bias} with the fragmentation function for gluons into charged hadrons\footnote{We thank A. Dumitru for pointing out an  error in this expression in an earlier version of the paper. After correcting the error, the resulting expression gives better agreement with data.},
\begin{equation}
\frac{d{\bar N}_{h}}{d^2p_{\bot}dy} = \int_{z_{\rm min.}}^1 \frac{dz}{z^2}\, \frac{d{\bar N}_{g}}{d^2q_{\bot}dy}\,D_{g\rightarrow h}\left(z=\frac{p_\perp}{q_\perp}, \mu^2\right) \,,
\label{eq:single-inclusive}
\end{equation}
where $D_{g\rightarrow h}(z,\mu^2)$ is chosen to be $6.05\,z^{-0.714}(1-z)^{2.92}$, corresponding to the LO parameter set of ~\cite{KKP}. The lower limit of the integral is determined from the kinematic requirement that $x_{1,2}\leq 1$. 
Our choice of fragmentation function is different from the fragmentation prescription of ref.~\cite{Lev-Rez} where the $p_\perp$ of gluons is scaled by $p_\perp/\langle z\rangle$, with $\langle z\rangle=0.5$. We also note that we do not have additional scales such as $p_{\perp, {\rm intrinsic}}$ in our comparison to data, in contrast to ref.~\cite{Lev-Rez}. 
\end{itemize}

\subsection{Results for the single inclusive hadron distribution}

Now that we have specified all the assumptions and parameters that go into eq.~\ref{eq:single-inclusive}, we are ready to present the results of our comparisons of the different saturation models with the data on rapidity and transverse momentum distributions for a wide range of collider energies up to the highest present LHC energy of $\sqrt{s}=7$ TeV and make projections for $\sqrt{s} =10, 14$ TeV.

\begin{figure}[h]
\centerline{
\includegraphics[width =7cm, height =7cm]{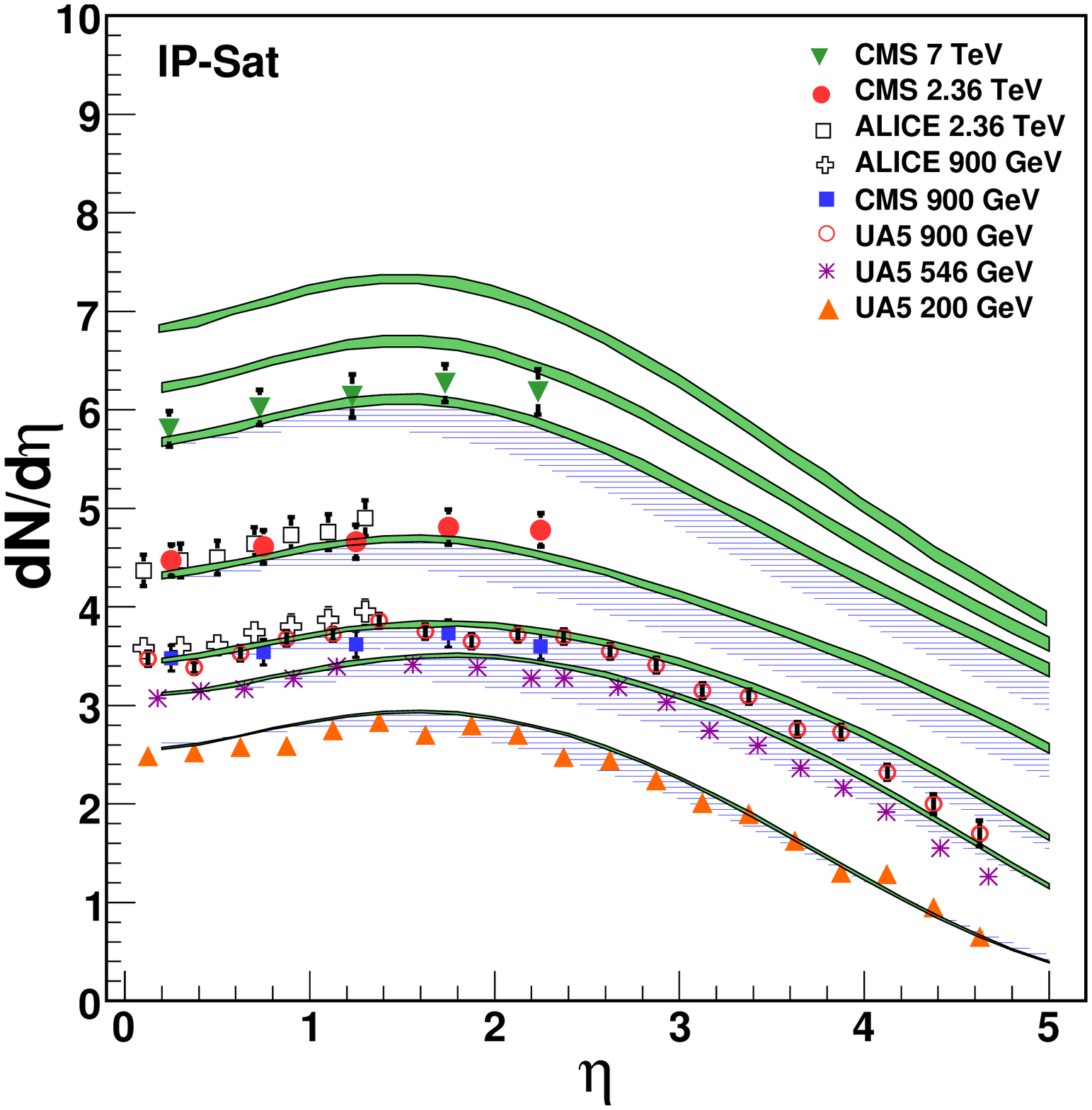}
\includegraphics[width =7cm, height =7cm]{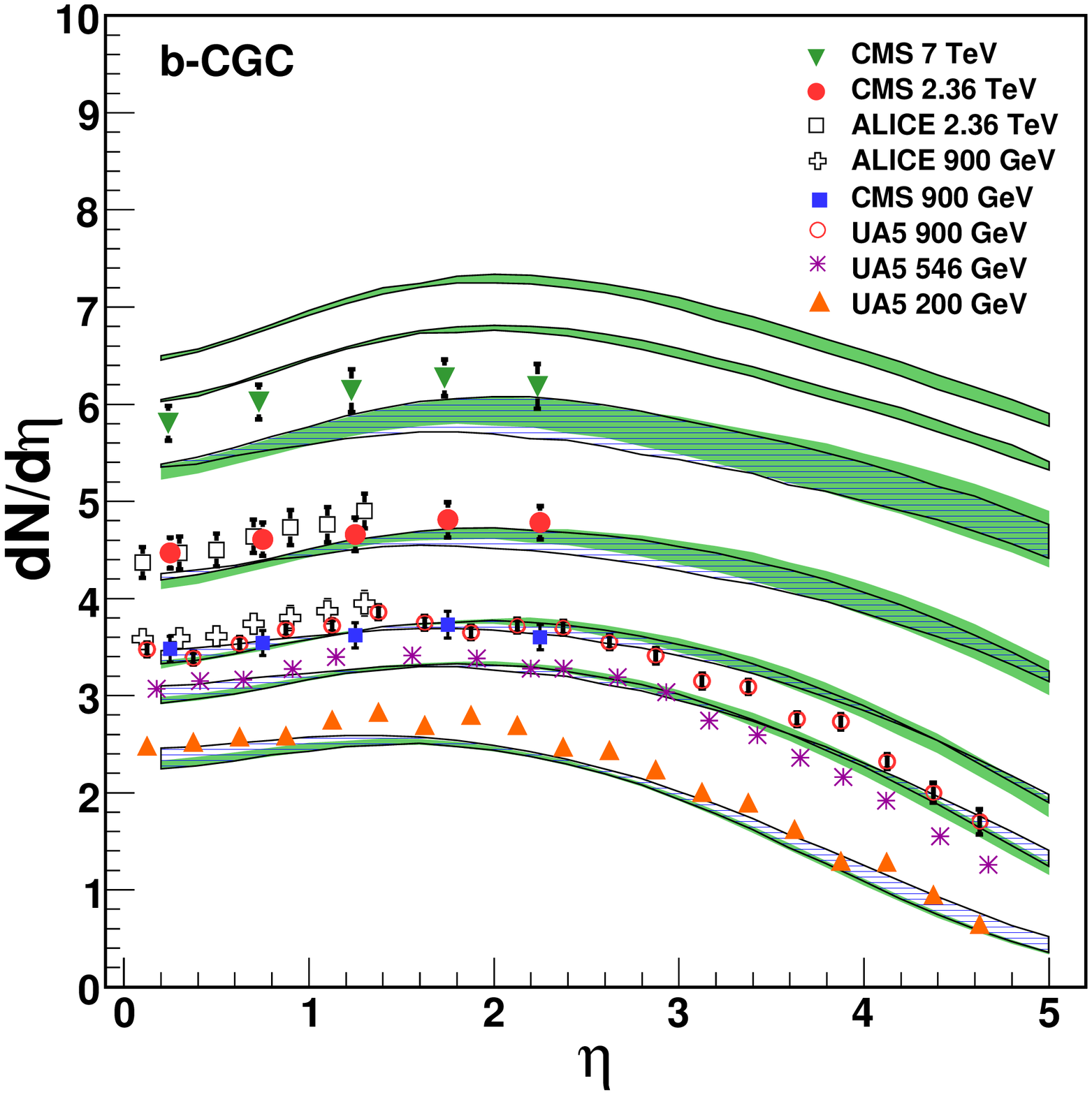}
}
\caption{dN/d$\eta$ obtained from comparing IP-Sat and b-CGC models to data from UA5~\cite{ua5}, ALICE~\cite{alice1} and CMS~\cite{cms}. The solid green band corresponds to uncertainties from different parameters;  the dashed band is due to the variation of the choice of mass term in the Jacobian relating $y$ to $\eta$. The two curves at the top in both panels correspond to projections in the two models for $\sqrt{s}=14$ and $10$ TeV respectively.}
\label{fig:fig_etadist}
\end{figure}
\begin{figure}
\centerline{
\includegraphics[width =7cm, height=7cm]{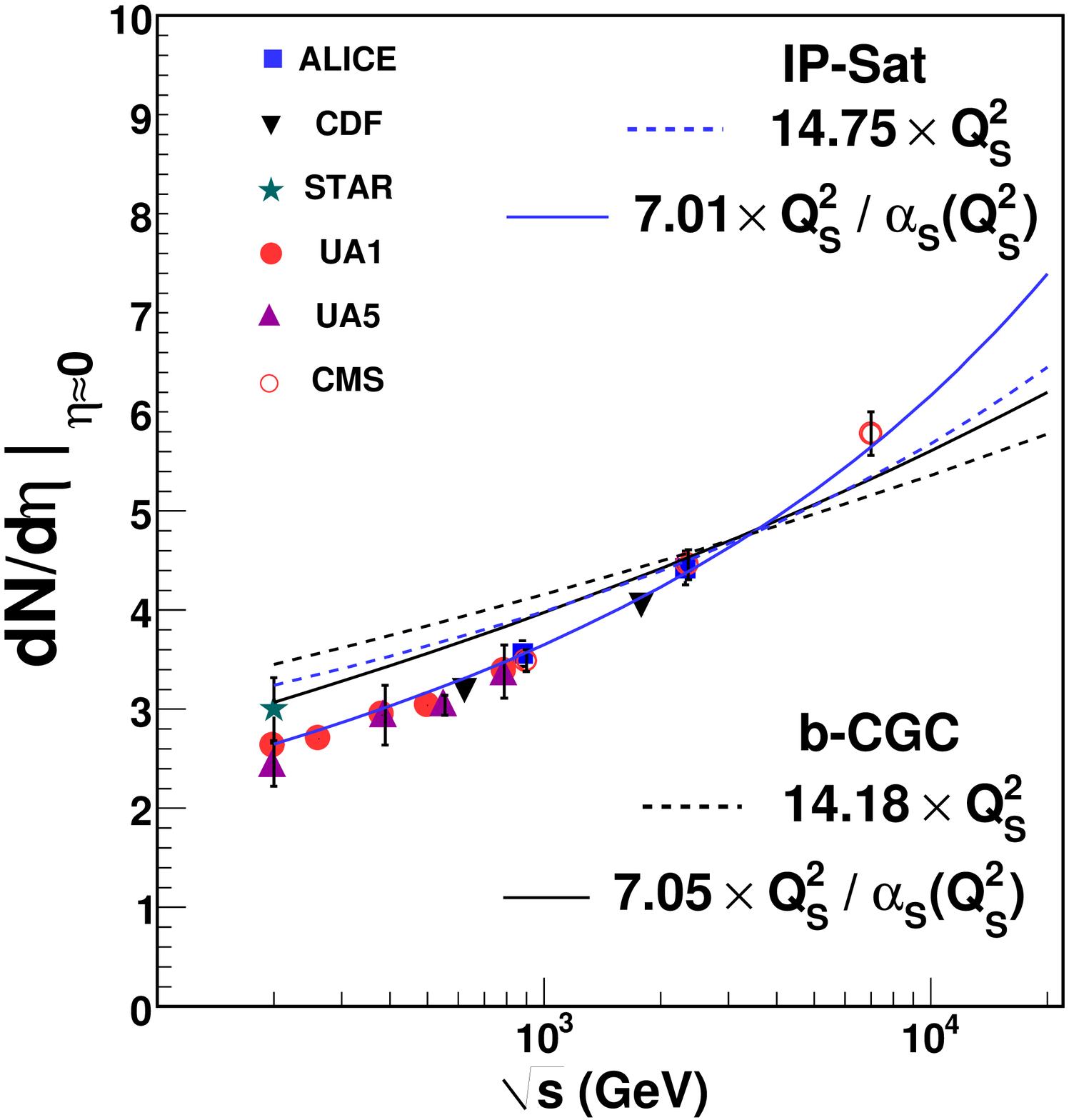}
\includegraphics[width =7cm,height =7cm]{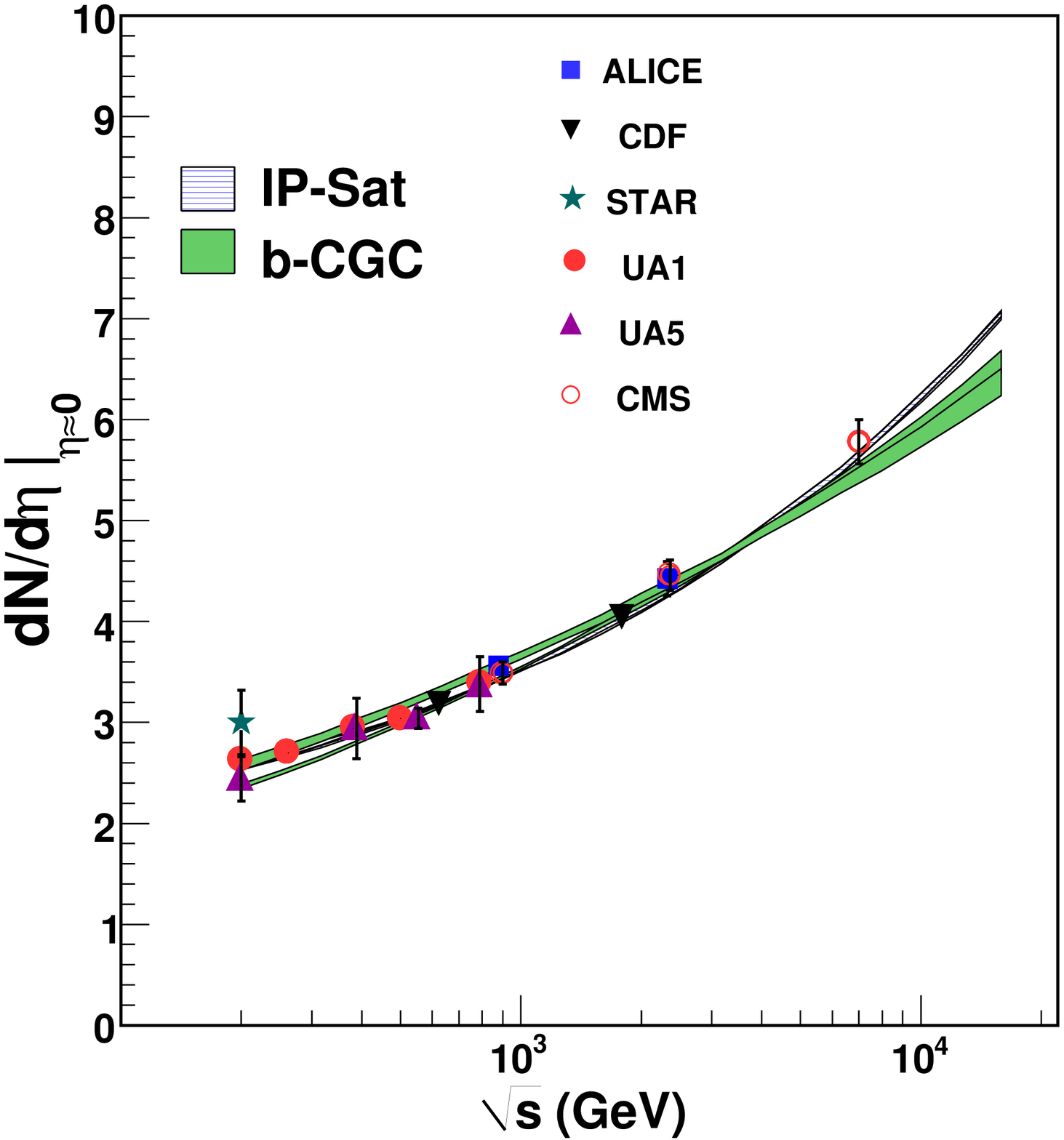}
}
\caption{ Average $\left.dN/d\eta\right|_{\eta\approx0}$ in the IP-Sat and b-CGC models. Left: Data plotted as a function of saturation scales (at median impact parameter $b=2$ GeV$^{-1}$) for both the models determined from the HERA data. Right: Average $dN/d\eta$ at $\eta=0$ from the $k_{\bot}$-factorized expression in eq.~\ref{eq:single-inclusive} from IP-Sat (solid green) and b-CGC (dashed) models. See text for further discussion. Experimental data points are from Ref.\cite{cms},~\cite{alice2},~\cite{ua1},~\cite{cdf},~\cite{star},~\cite{ua52}.}
\label{fig_avgeta}
\end{figure}

Fig.~\ref{fig:fig_etadist} shows the pseudo-rapidity distribution obtained by integrating eq.~\ref{eq:ktfact1} over $p_{\bot}$ for different mass terms in the Jacobian and parameters given in the tables~\ref{tab_ipsat} and \ref{tab_bcgc} for the IP-Sat and b-CGC dipole models respectively. The uncertainties corresponding to the choices of parametrizations and the infrared mass scale are shown in bands. In the IP-Sat model, the normalization is performed to the $\sqrt{s}$ dependence of $dN/d\eta$ at $\eta=0$ as shown in fig.~\ref{fig_avgeta}--this fixes the two parameters in $b_{\rm max.}$ we discussed previously. In the b-CGC model, because there is one less parameter (the coefficient of $\ln{s}$ in $b_{\rm max.}$ is zero), the 
normalization can be performed to the rapidity distribution at one fixed energy. In the figure shown, the normalization is performed for $\sqrt{s}=900$ GeV; choosing a lower energy 
corresponds to a $< 10\%$ uncertainty in the overall normalization.

We didn't use a fragmentation function in computing the pseudo-rapidity distributions because the rapidity distribution is vastly dominated by contributions below $p_\perp =1$ GeV, where fragmentation functions are likely not reliable. We have varied the mass term in the Jacobian corresponding to eq.~\ref{eq:Jacobian} in the range $0.2$--$0.4$ GeV, corresponding to an infrared scale of order $\Lambda_{QCD}$.  (In each case we chose this mass term to be equal to the one we use to regulate the infrared divergence of the unintegrated gluon distribution in eq.~\ref{eq:unint-gluon}). The effect of the extrapolation parameter $\lambda_0$ in eq.~\ref{eq:largex} is significant only at lower energies and and higher values of $\eta$. The agreement with data of both models is quite good with the IP-Sat model providing a somewhat better agreement at the highest energies. 

\begin{figure}[h]
\centerline{
\includegraphics[width =7cm,height =7cm]{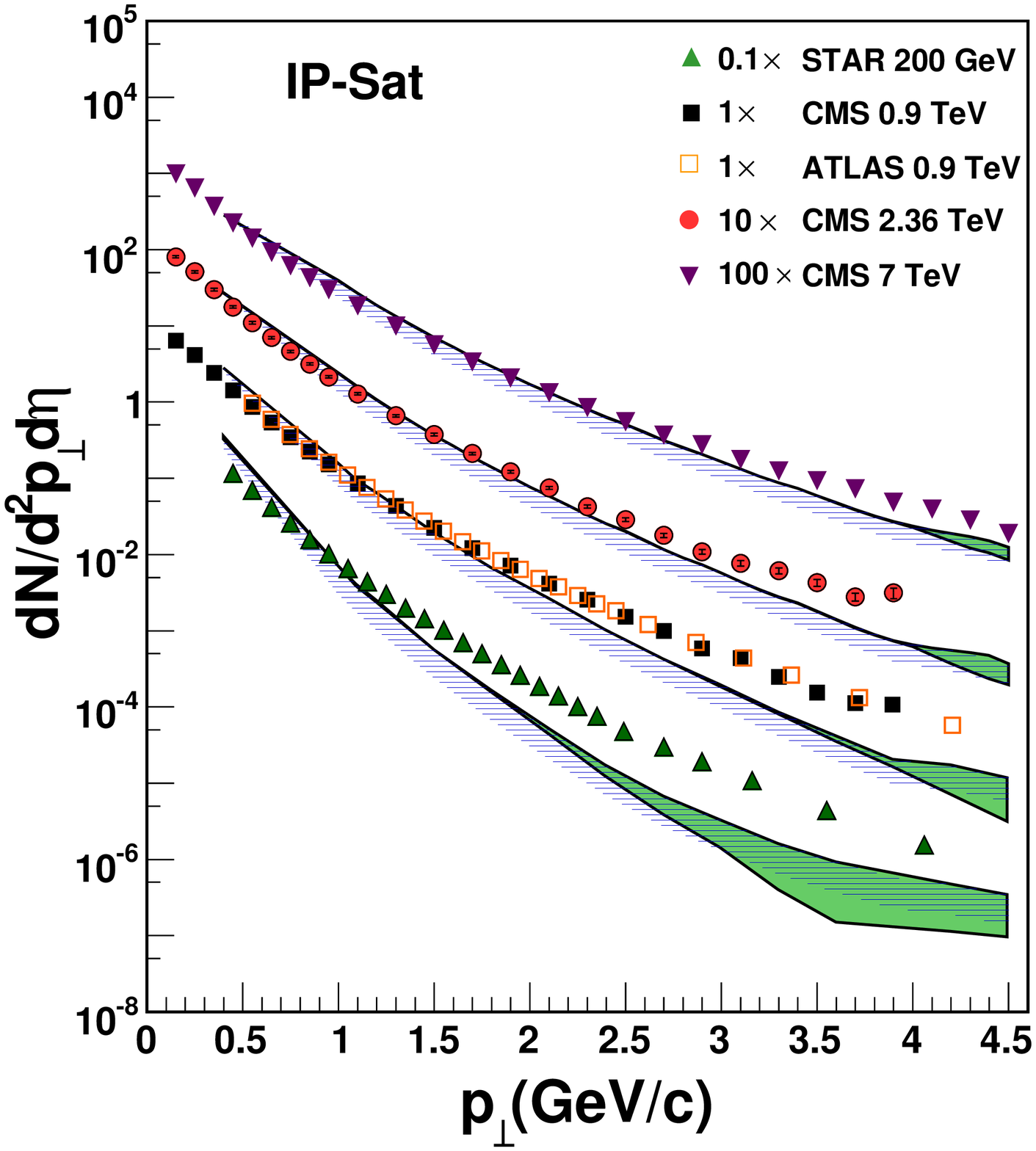}
\includegraphics[width =7cm,height =7cm]{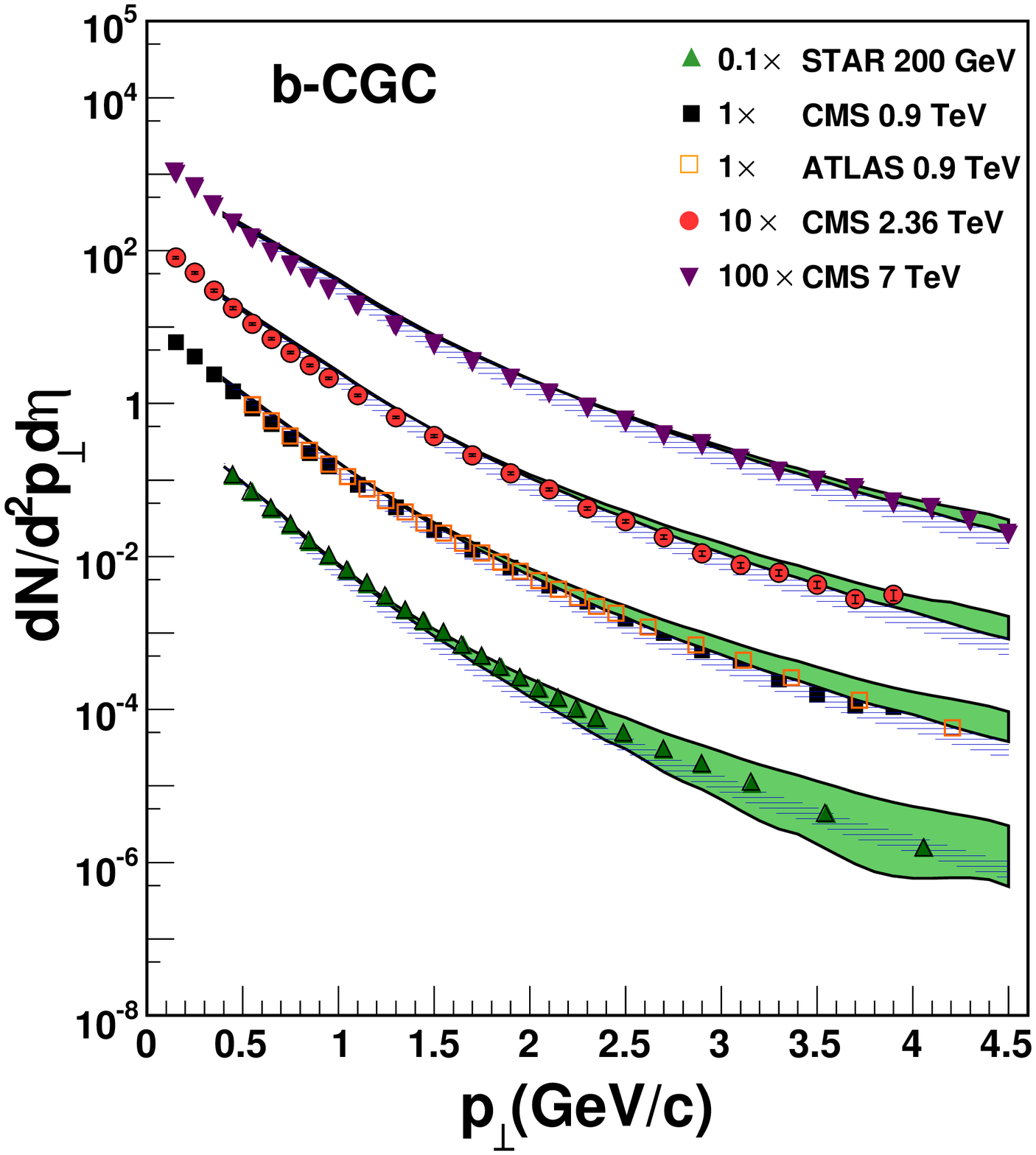}
}
\caption{dN/d$^2p_{\bot}$d$\eta$ in the IP-Sat and b-CGC models. The solid green band corresponds to uncertainties from different parameters and the dashed band is due variation of the mass term in the Jacobian relating $\eta$ to $y$. The $p_{\bot}$ distribution is averaged over the $\eta$ range of $\pm$2.4. The experimental data points are from CMS~\cite{cms}, STAR~\cite{star} and ATLAS~\cite{atlas}}
\label{fig:fig_ptdist}
\end{figure}

In fig.~\ref{fig:fig_ptdist}, the corresponding $p_\perp$ distributions, with the previously specified fragmentation prescription, is shown. The IP-Sat model shows a poor agreement  with data for the lower energies at high $p_\perp$, as does the b-CGC model with some of the HERA parameter sets. At higher $p_\perp$, the results are sensitive to physics at $x \geq 0.01$, which is parametrized very simply in the models. On the reverse side of the coin, one should anticipate a better agreement in the same $p_\perp$ window at higher energies. Indeed, a systematically better agreement is seen in both models with the data at the higher energies for all parameter sets.

Fig.~\ref{fig_avgeta} shows the average value of $dN/d\eta$ calculated at $\eta$=0. The left plot of fig.~\ref{fig_avgeta} shows a fit of $dN/d\eta$ using different functional forms for the saturation scales. We considered both $Q_S^2$ and $Q_S^2/\alpha_S(Q_S)$. In the CGC framework, one would expect the latter. However, if the running is not significant in the energy range of interest, the former is also possible. Indeed, a good fit to the CMS data was obtained~\cite{Larry-Michal} with a simple form of the saturation scale {\it a la} the Golec-Biernat model~\cite{GolecW}.  In our case, the comparison is made within the framework of the IP-Sat and b-CGC models which give better fits to the HERA data and are sensitive to the impact parameter profile of the gluon distribution in the proton. For the comparison with the LHC data in the left panel of fig.~\ref{fig_avgeta}, we use the value of $Q_S$ at the median value of $s_{\bot}$=2 GeV$^{-1}$. The dependence of $dN/d\eta$ on the purely $Q_S^2$ functional form is not very good, while the $Q_S^2/\alpha_S(Q_S)$ form does much better for the IP-Sat model. For the running of $\alpha_S$,  we chose $Q_S^2(s_{\bot})$ at $s_{\bot}$=0 to restrict its running to $\alpha_S$ below 0.5. Fig.~\ref{fig_avgeta} (right panel) shows by way of comparison, a comparison of $dN/d\eta$ at $\eta=0$  as a function of $\sqrt{s}$ to IP-Sat and b-CGC models. In the IP-Sat model, a good fit is ensured because a fit to this energy dependence is what determines the parameters of $b_{\rm max.}$--see fig.~\ref{fig_sigmain} and related discussion. In the b-CGC model, the curve is a prediction and is seen to be a very good fit to the data. 

\begin{figure}
\centerline{
\includegraphics[width =7cm,height =7cm]{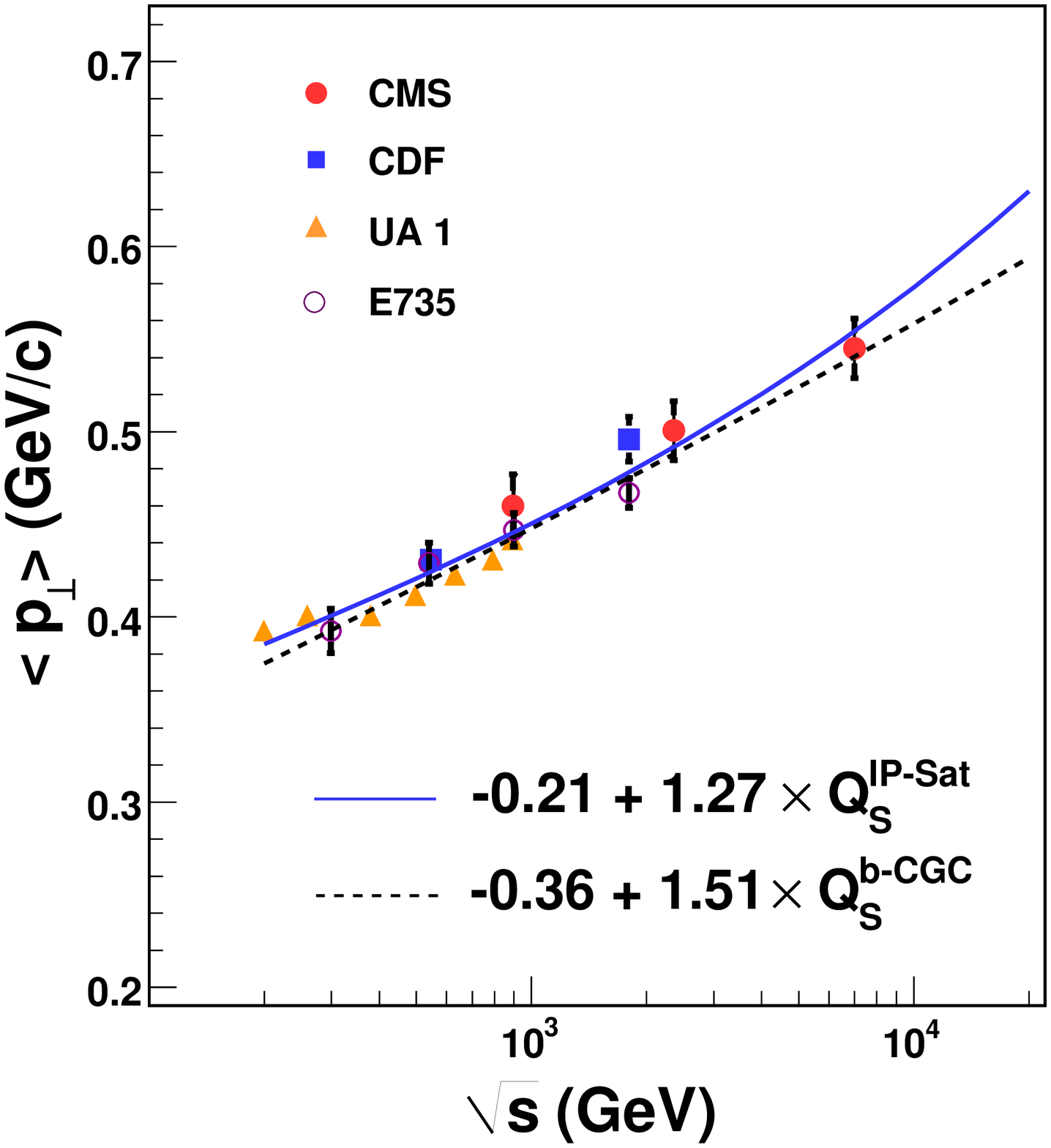}
\includegraphics[width =7cm,height =7cm]{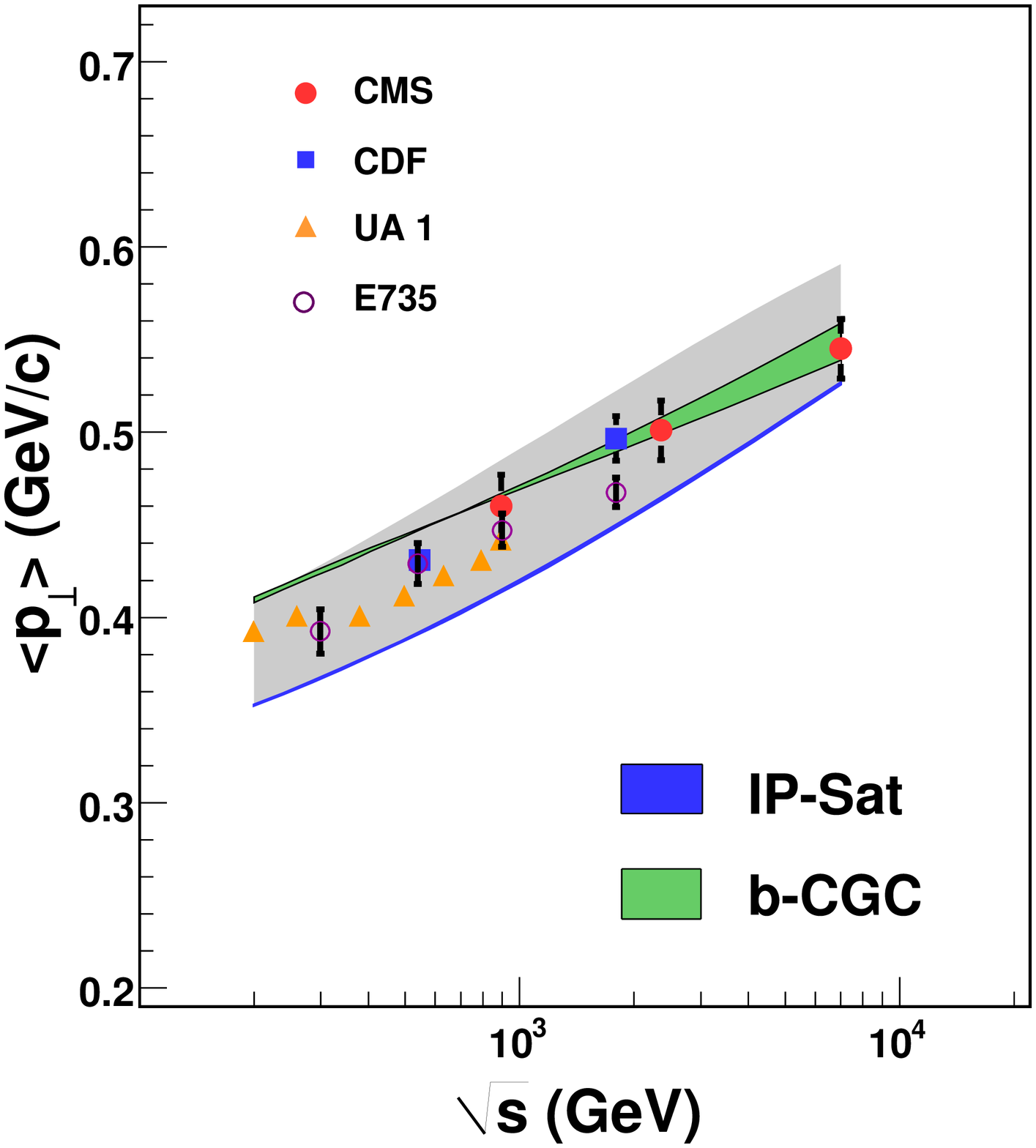}
}
\caption{ Average $p_{\bot}$ obtained from IP-Sat and b-CGC models compared to data. Left: Function of saturation scales fitted for both the models. Right: Average $p_{\bot}$ from the $k_{\bot}$-factorized expression in eq.~\ref{eq:single-inclusive}. The thin (colored) bands correspond to uncertainties arising from different parameters in table~\ref{tab_ipsat} and table~\ref{tab_bcgc} and mass term=$0.2$ GeV. The thick (gray) band shows the sensitivity to variation of mass term in the range $0.2$--$0.3$ GeV (with lower mass corresponding to lower $\langle p_{\bot} \rangle$ ). Experimental data points are from Ref.~\cite{cms}~\cite{e735}~\cite{cdf}~\cite{ua1}.}
\label{fig_avgpt}
\end{figure}

The energy dependence of $\langle p_{\bot}\rangle$ is shown in fig.~\ref{fig_avgpt}.  In the left panel, it is shown that in this case one obtains a good linear dependence of $\langle p_{\bot}\rangle$ on $Q_S$ in both the IP-Sat and b-CGC models as the c.m energy is varied. This is as seen previously~\cite{Larry-Michal}. The right plot shows $\langle p_{\bot}\rangle$ versus $\sqrt{s}$ computed in the IP-Sat and b-CGC models. Here one sees that the results are quite sensitive to choice of the infrared cut-off. 

\begin{figure}[h]
\centerline{
\includegraphics[width =7cm, height=7cm]{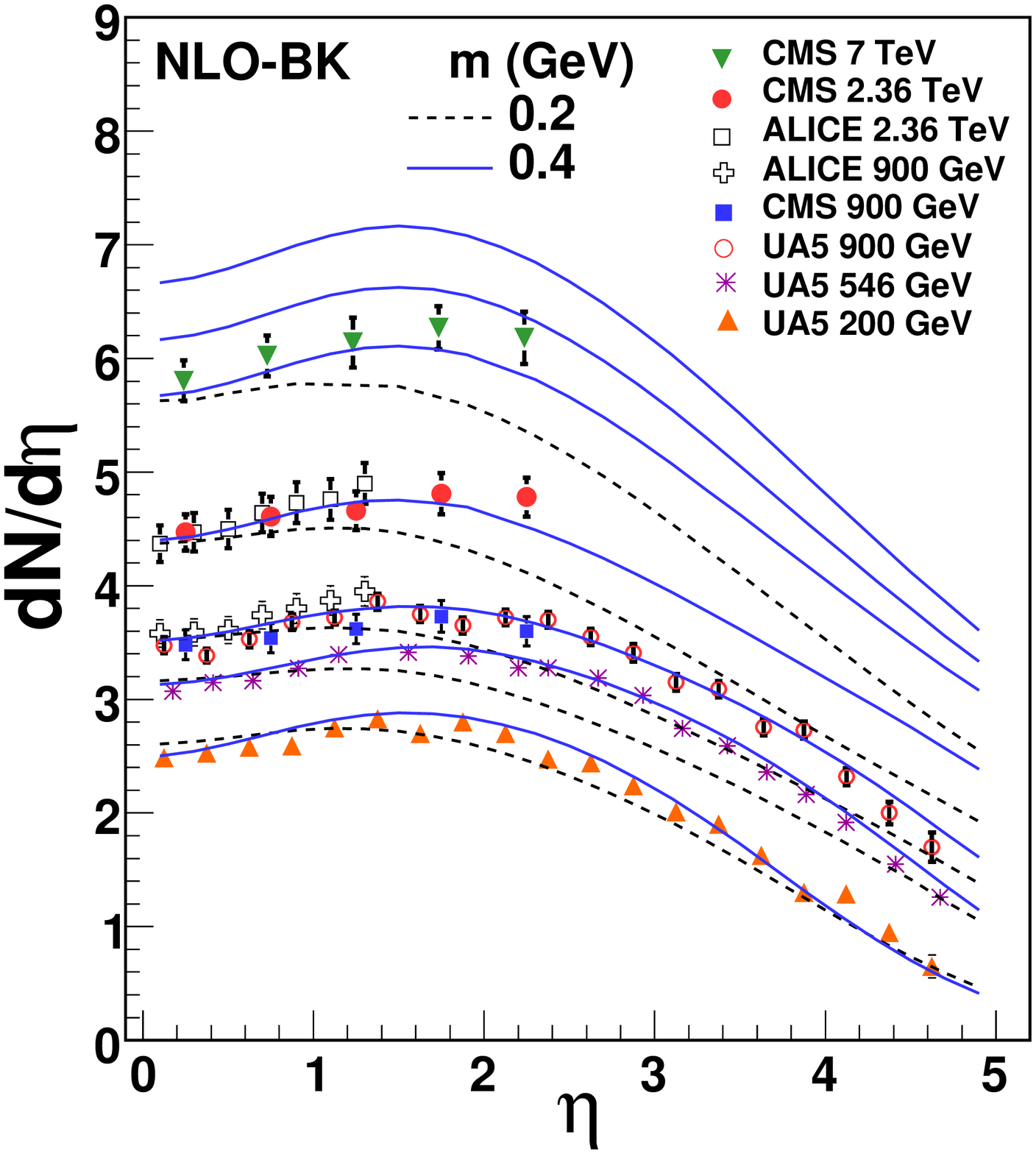}
\includegraphics[width =7cm, height=7cm]{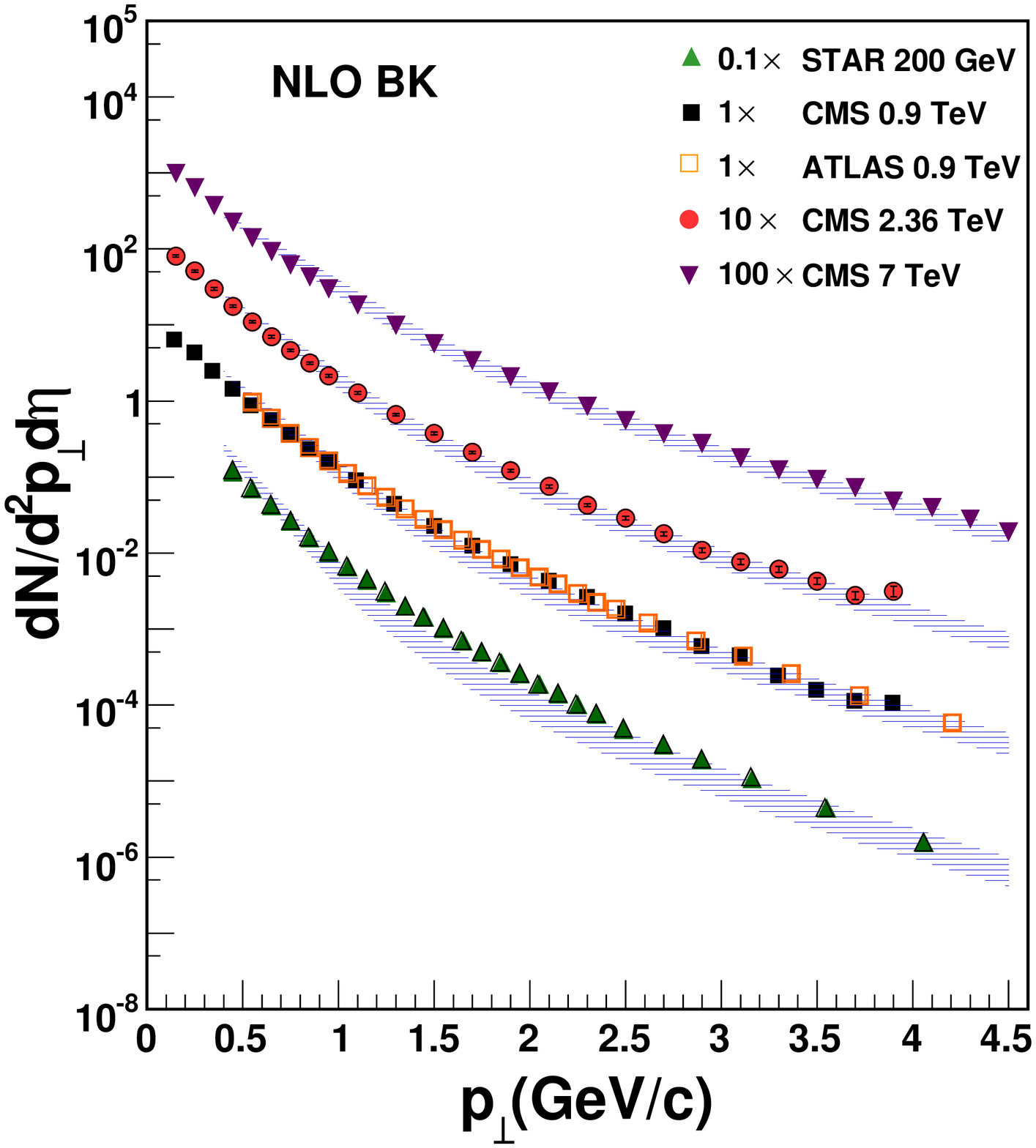}
}
\caption{Pseudo-rapidity and $p_{\bot}$ distribution in the NLO-BK model compared to data. The uppermost two plots in the  left panel correspond to predictions for $\sqrt{s}= 14, 10$ TeV with $m=0.4$ GeV. The $p_{\bot}$ distribution is averaged over the $\eta$ range of $\pm $2.4. The band corresponds to the variation  $m= 0.2$--$0.4$ GeV in the Jacobian relating $\eta$ to $y$. }
\label{fig_nlobk}
\end{figure}
Fig. \ref{fig_nlobk} shows the $\eta$ and $p_\perp$ distributions computed in the NLO-BK model. Only MV initial conditions are considered. In this model, the impact parameter dependence of the inclusive gluon multiplicity is given by 
\begin{equation}
 {\cal T} (b) = {1\over \pi b_{\rm max.}^2}\, \Theta\left(b_{\rm max.}-b\right) \, . \nonumber
\label{eq_tbnlobk}
\end{equation}
 Therefore $d\bar{N_g}(b)/d\eta = d\bar{N_g}/d\eta ~{\cal T} (b)$, where $b_{\rm max.}$ is a parameter that can be absorbed in the normalization. If no dependence of $b_{\rm max.}$ on $\sqrt{s}$ is assumed, the model considerably overestimates the single inclusive data at LHC energies. This is likely a consequence of the fact that the NLO-BK doesn't take into account the impact parameter dependence of the saturation scale. In this case, the  values for the inclusive gluon multiplicity correspond to the values for the zero impact parameter, which is considerably higher than the minimum bias values. The agreement with data is improved considerably by allowing $b_{\rm max.}$ to depend on $\sqrt{s}$. The denominator $\pi b_{\rm max.}^2$ therefore provides an energy dependent normalization. The same approach discussed previously for the IP-Sat model is employed to extract the $\sqrt{s}$ dependence of $b_{\rm max.}$ chosen to be of the form $b_{\rm max.} = b_0 + C\,\ln(s)$. From a fit to the average $dN/d\eta$ as a function of energy, one obtains $b_0 \sim 5.6-7.55$ GeV$^{-1}$ and $C \sim 0.23-0.46$ depending on the choice of infrared cut-off.

\section{Multiplicity distribution}

There are several sources of multiplicity fluctuations in high energy hadronic collisions. These can arise from fluctuations in the number of wee partons, in their distribution with impact parameter and their distribution in rapidity~\cite{MiettPump}. In this paper, we will consider particle production in a relatively small rapidity window (parametrically of order $\Delta \eta \leq 1/\alpha_S$), so fluctuations in rapidity will not be an important source of fluctuations. Let us first consider fluctuations in multiplicity for a 
fixed impact parameter. In this case, the CGC framework allows for a systematic treatment of inclusive multi-particle production~\cite{GelisLV} in the Glasma~\cite{LappiM}. The largest contribution to multi-particle production comes from 
diagrams that appear superficially disconnected, but are connected by averaging over color correlations in an event and over all events. This is the formal basis of the Glasma flux tube picture~\cite{DumitruGMV}, and was previously used to compute backward-forward correlations~\cite{ArmestoMP,LappiM1}, two particle correlations~\cite{DumitruGMV,DuslingGLV}, three particle correlations~\cite{DuslingFV}, and n-particle correlations~\cite{GelisLM}. It is this last computation that will concern us here. The n-particle correlations obtained by averaging over color sources in the CGC picture are those that would be generated by the negative 
binomial distribution~\cite{GelisLM}. To the best of our knowledge, this derivation of the negative binomial distribution as arising from particle production from Glasma flux tubes is the first such {\it ab initio} derivation in a QCD framework~\footnote{The negative binomial distribution has of course been known for a long time~\cite{Dremin-Wolf} to provide good fits to collider p+p data. More recently shown to describe the multiplicity distributions in A+A collisions at RHIC~\cite{phenix}. However, a microscopic derivation of this distribution from QCD was previously lacking.}

The negative binomial distribution is given by
\begin{equation}
P_n^{\rm NB}({\bar n}, k) = \frac{\Gamma(k+n)}{\Gamma(k)\Gamma(n+1)}\,\frac{{\bar n}^n k^k}{({\bar n} +k)^{n+k}} \, .
\label{eq:NB}
\end{equation}
The distribution is characterized by two parameters, the mean multiplicity ${\bar n}$ and the parameter $k$. 
As is well known, in the limit $k\rightarrow \infty$, this distribution reduces to the Poisson distribution. In the limit $k\rightarrow 1$, one obtains the 
Bose-Einstein distribution. The variance of the distribution is given by $\sigma^2 = {\bar {n^2}} - {\bar n}^2 = {\bar n} + {\bar n}^2/k$. In the Glasma flux tube approach, $k$ is not an arbitrary parameter; instead, it is computed to be 
\begin{equation}
k = \zeta {(N_c^2-1) Q_S^2 S_\perp \over 2\pi} \, ,
\label{eq:flux1}
\end{equation}
where $\zeta$ is a dimensionless non-perturbative parameter which will be discussed shortly and $S_\perp$ is the overlap area of the two hadrons. 

As we mentioned previously, the negative binomial distribution is obtained at a fixed impact parameter, so ${\bar n} \equiv {\bar n}(b)$ and $k\equiv k(b)$. In particular, the latter parameter must be interpreted as being proportional to the number of flux tubes (or interacting ``hot spots") $S_\perp /1/Q_S^2$ at a given impact parameter. Because $Q_S^2$ grows with energy, $k$ has a very particular energy dependence. Here, the parameter $k$ in eq.~\ref{eq:flux1} is determined as follows. For a given impact parameter, we define 
\begin{equation}
Q_S^2\, S_\perp = \int d^2 x_\perp Q_S^2 (x_\perp) \, , \nonumber
\end{equation}
where the integral on the r.h.s is performed over the overlap area of the two protons at a given parameter. At each given transverse position $x_\perp$ in the overlap area, we choose $Q_S(x_\perp) = {\rm min.}\,\{Q_S^A, Q_S^B\}$, where $Q_S^A$ and $Q_S^B$ are respectively the saturation scales of the two colliding protons at that $x_\perp$. This choice is motivated by the fact that the inclusive multiplicity of produced gluons is much more sensitive to the smaller of the two saturation scales~\cite{DumitruM}. 
\begin{figure}[h]
\centerline{
\includegraphics[width =6cm, height=6cm]{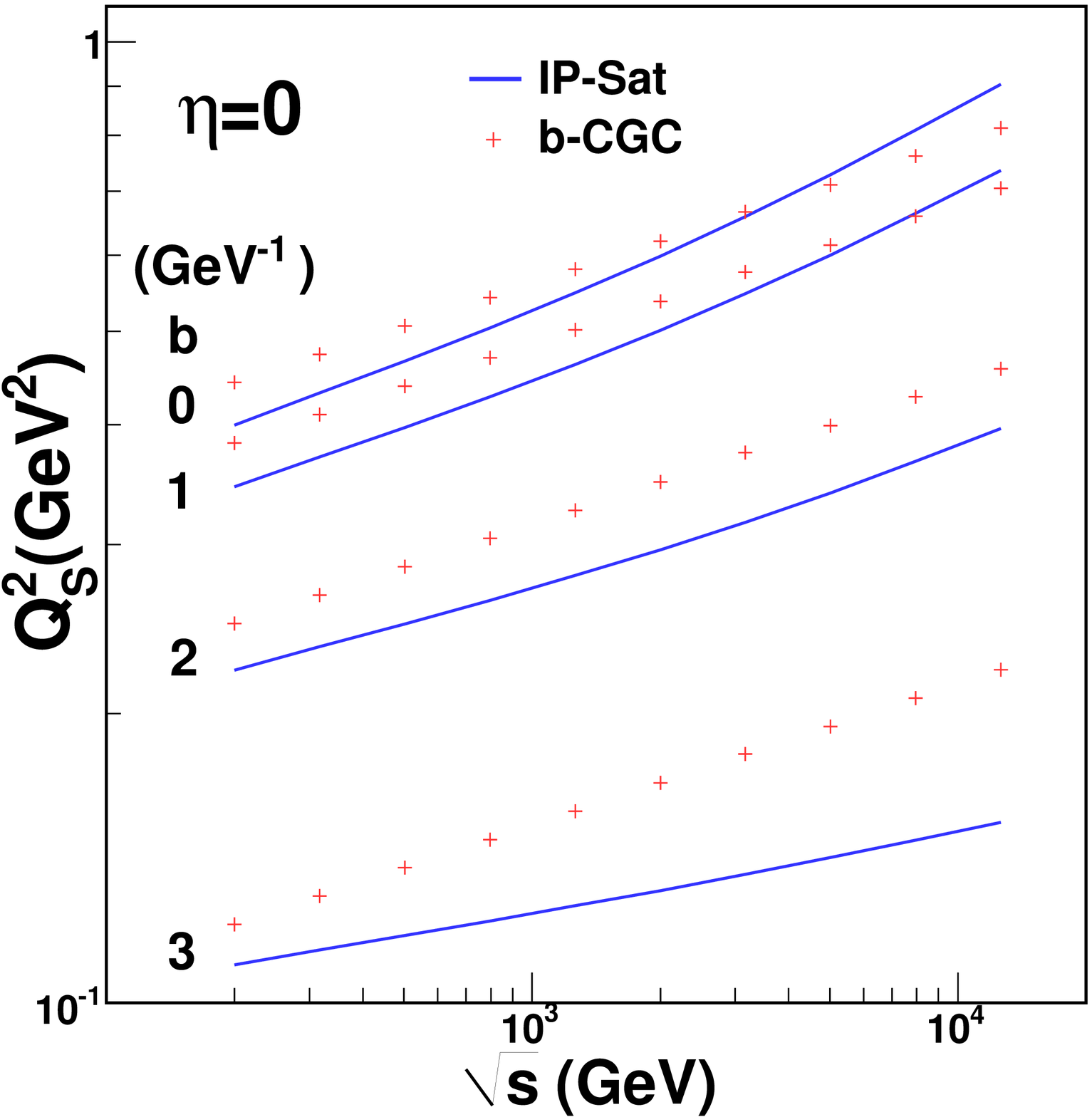}
\includegraphics[width =6cm, height=6cm]{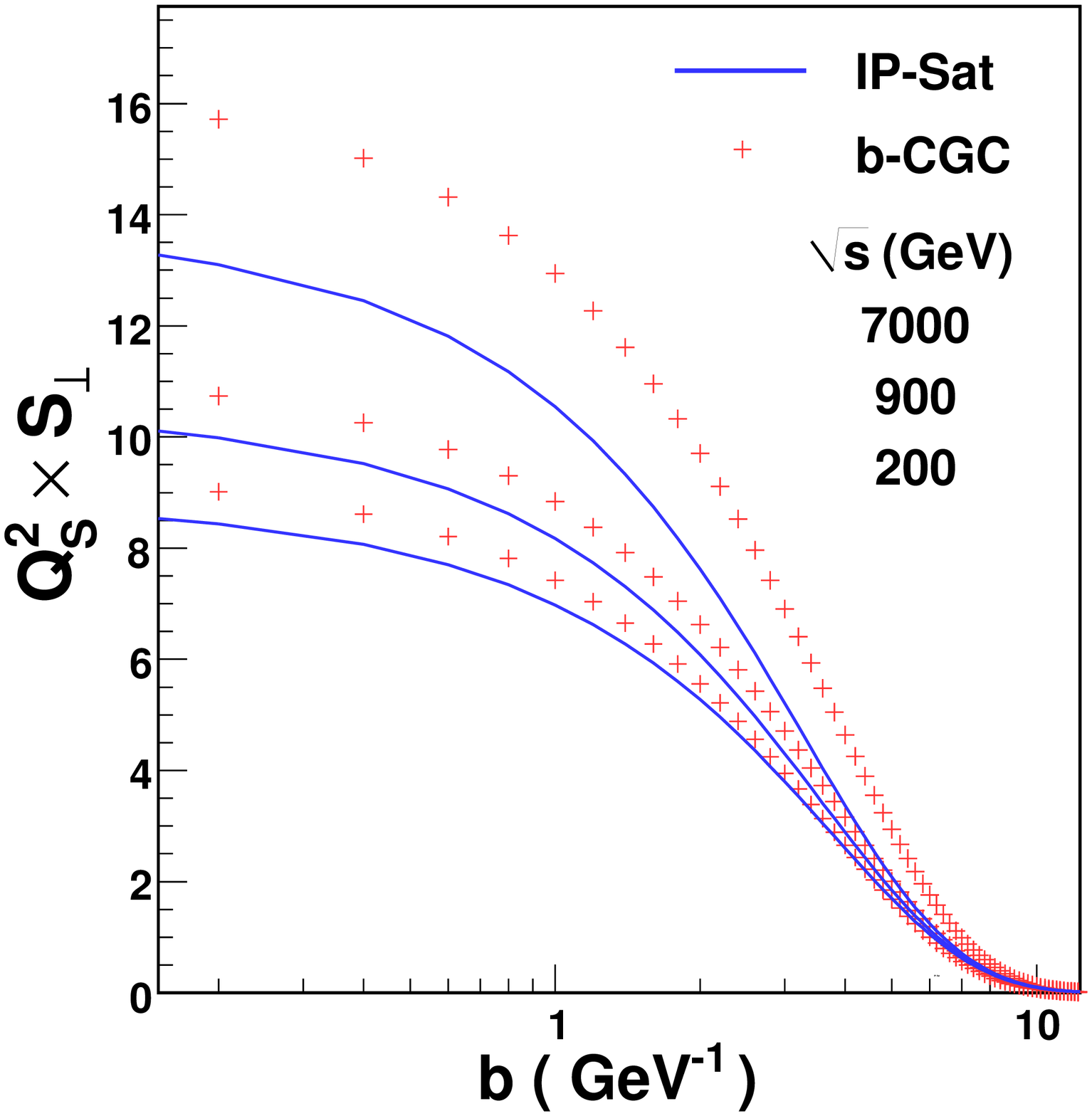}
}
\caption{Left: Variation of the saturation scale with the collision energy. Right: Variation of $Q_{S}^2 S_{\bot}$ for different impact parameters and the c.m. energy of the collision. The solid lines are for the IP-Sat model and crosses for the b-CGC model, in each case for the parameters providing the best fit to the HERA data.}
\label{flux_tubes}
\end{figure}
In fig.~\ref{flux_tubes} (left) we plot the saturation scale as a function of $\sqrt{s}$ for different impact parameters. The right plot has on the $y$-axis the quantity that appears in 
eq.~\ref{eq:flux1} determined by the procedure we described. We observe that a stronger dependence of this quantity is seen for the b-CGC model. 
With the exception of the parameter $\zeta$, we have everything necessary to compute eq.~\ref{eq:NB} at a given impact parameter. 

Fluctuations in impact parameter are treated as follows. The overlap function for two protons at a given impact parameter (see fig.~\ref{fig:ppcol} (right)) can be expressed as 
\begin{equation}
T_{\rm pp}(\bt) = \int d^2 {\mathbf s}_\perp T_{\rm p}({\mathbf s}_\perp) \, T_{\rm p}({\mathbf s}_\perp - \bt)
\end{equation}
where $T_{\rm p}$ is given for instance in the IP-Sat model by eq.~\ref{eq:IPsat-imp-par}. Our knowledge of the HERA diffractive data therefore allows one to compute $T_{\rm pp}$ in the saturation models. 
The probability distribution for an inelastic collision at a given impact parameter is given in impact parameter eikonal models as~\cite{D'Enterria,FSW,Steinberg}
\begin{equation}
{dP_{\rm inel.}^{\rm eik.}\over d^2 \bt} = {1-\exp\left(-\sigma_{\rm gg} T_{\rm pp}\right) \over \int d^2 \bt \left(1-\exp\left(-\sigma_{\rm gg} T_{\rm pp}\right)\right)} \, .
\label{eq:imp-prob-eik}
\end{equation}
In general, $\sigma_{\rm gg}$ is an energy dependent quantity estimated to be the elementary cross-section for gluon-gluon scattering. 
Alternately, in our framework, an estimate for this quantity in our framework involving no additional parameters is
\begin{equation}
{dP_{\rm inel.}^{\rm dip.}\over d^2 \bt} = {\frac{dN_g}{dy} (\bt) \over \int d^2\bt \frac{dN_g}{dy}(\bt)} \, .
\label{eq:imp-prob-dip}
\end{equation}
\begin{figure}[h]
\includegraphics[width =6cm, height=6cm]{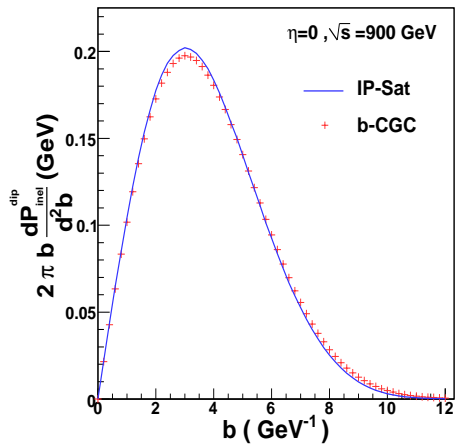}
\caption{Probability distribution as a function of impact parameter for an inelastic collision computed using eq.~\ref{eq:imp-prob-dip} for the b-CGC and IP-Sat models.}
\label{fig_imp_prob_dist}
\end{figure}
This expression of course gives unity when both sides are integrated over impact parameter. The result for $2 \pi b ~dP_{\rm inel.}^{\rm dip.}/d^2 \bt$ at a fixed $\sqrt{s}=900$ GeV is shown in fig.\ref{fig_imp_prob_dist} and is a sharply peaked distribution at $\sim 3$ GeV$^{-1}$. The distribution plotted is insensitive to $\sqrt{s}$ and to the infrared scale $m$. We should note here that the inclusive multiplicity at a given impact parameter is given by ${\bar n}(\bt) = P_1 + 2\,P_2 + 3\, P_3+\cdots$, while the inelastic probability is given by $P_{\rm inel.}= P_1 + P_2 + P_3+\cdots$.  In general, the two are of course not the same, so using eq.~\ref{eq:imp-prob-dip} to estimate the inelastic probability is strictly not correct. Our justification here is that eq.~\ref{eq:imp-prob-dip} has the right qualitative behavior and involves one less parameter than eq.~\ref{eq:imp-prob-eik}. It is very desirable to obtain a better estimate for eq.~\ref{eq:imp-prob-dip}. For a detailed discussion of the problem of computing  multi-particle production in the presence of strong time dependent sources, see ref.~\cite{GelisV}. 

\begin{figure}[t]
\includegraphics[width =7cm, height=7cm]{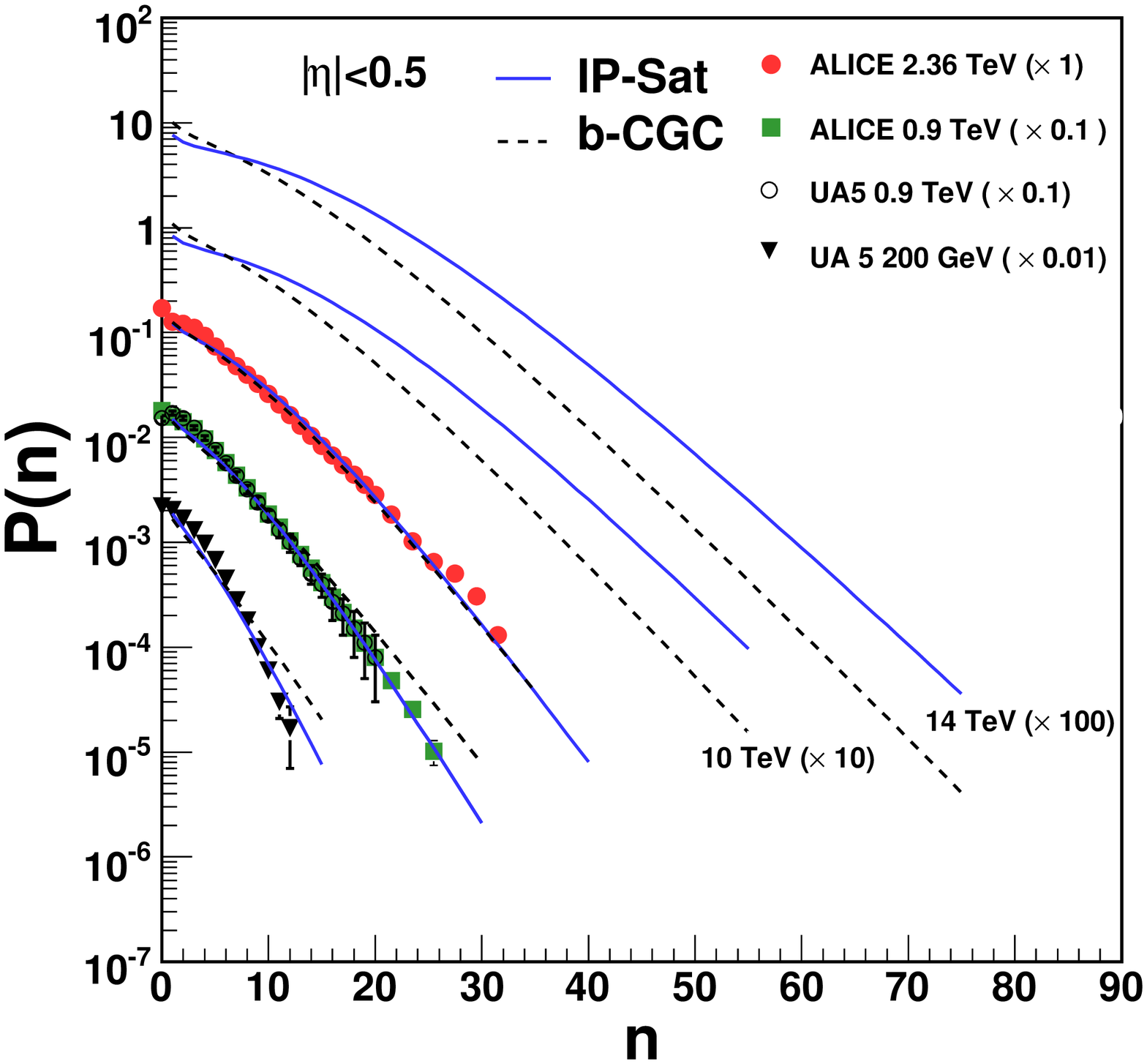}
\includegraphics[width =7cm, height=7cm]{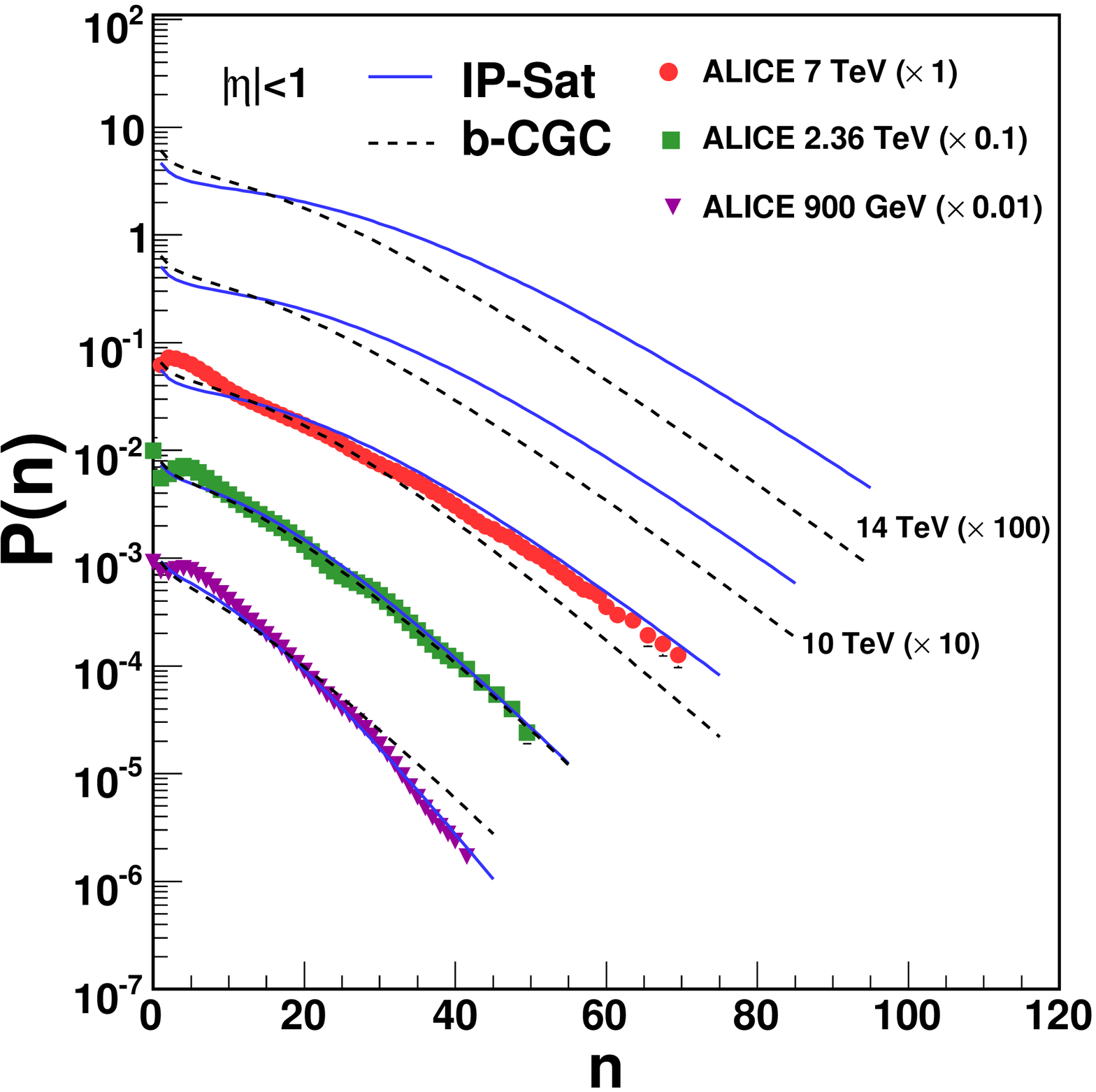}
\caption{Probability distribution of the gluon multiplicity computed in the IP-Sat model (parameter set I) and the b-CGC model (parameter set II) compared to the UA5\cite{ua53} and ALICE\cite{alice2} data for different $\eta$ ranges. Left: Multiplicity distribution for the $\eta$ range $\pm 0.5$. Right: Multiplicity distribution for the $\eta$ range $\pm 1$.} 
\label{fig_multdist}
\end{figure}

With the stated assumptions (and caveats) we are now in a position to compute the probability distribution as a function of multiplicity.  By convolving the probability distribution for producing $n$ particles at a given impact parameter (eq.~\ref{eq:NB}) with the probability for an inelastic collision at that impact parameter (from eq.~\ref{eq:imp-prob-dip}), one obtains the expression 
\begin{equation}
P(n) = \int d^2 \bt {dP_{\rm inel.} \over d^2\bt}\, P_n^{\rm NB}({\bar n}(\bt), k(\bt))
\label{eq:mult-dist.}
\end{equation}
The results for this quantity are shown in fig.~\ref{fig_multdist}.  Note that since the input here is the average inclusive multiplicity at a given impact parameter (as opposed to the minimum bias inclusive multiplicity that was compared to data), this quantity needs to be normalized as well.  We do so by fitting the multiplicity distribution corresponding to the lowest energy UA5 data set at $\sqrt{s} = 200$ GeV for the normalization of $\bar n$. In the b-CGC model, a fit of the overall normalization to the single inclusive minimum bias distribution at a given energy gives the same value ( to an accuracy of $<$ 5 \%) as that obtained for the same quantity if fit to the multiplicity distribution instead. In the IP-Sat model, one obtains the same normalization constant if 
$b_{\rm max.}=2\, b_{\rm rms.}$. If one chooses the form $b_{\rm max.} = b_0 + C\ln(s)$ we described previously, there is a 60\% discrepancy between the two choices of fixing the normalization. The agreement between data and model is remarkably good for IP-Sat distribution for all energies and rapidity cuts\footnote{The results are insensitive to the choice of the infrared cut-off $m$.}. For the b-CGC model, the 
agreement with data for $|\eta| < 0.5$ is quite good for the 2.36 TeV ALICE data~\cite{alice2} but shows deviations for other energies at the highest multiplicities\footnote{Our computation of the saturation scale is performed for $\eta=0$ while the data we compare with are averaged over $|\eta| < 0.5$ and $|\eta| < 1$. This discrepancy may lead to small corrections to our results.}. 

An important point regarding the comparison of the models to data in fig.~\ref{fig_multdist} concerns the magnitude of the parameter $\zeta$ in eq.~\ref{eq:flux1}  which is fit to the data. In the b-CGC model, it is extracted to be $0.25$ and it is $0.35$ in the IP-Sat model~\footnote{We have not attempted any fine tuning of our fits at this stage. A good agreement can be obtained for the 2.36 TeV and 7 TeV data in the b-CGC model by decreasing $\zeta$ by 
20\% but this gives a significantly worse fit at lower energies. However, because the model's validity is questionable at lower energies, and because of other sources of uncertainty we have articulated, the b-CGC model still provides a ``competitive" mechanism for multi-particle production.}. This quantity was recently computed non-perturbatively {\it ab initio} by 
solving the Yang-Mills equations numerically~\cite{LappiSV} for two gluon correlations from gauge fields generated in the collision of two dense color sources~\cite{LappiSV}. The results of the numerical computation vary depending on parameter choices in the range $\zeta\sim 0.3$-$1.5$--the essential point is that the result is a number of order unity. It is very encouraging that the values extracted from the data for $\zeta$ in the saturation models are numbers of this order~\footnote{Nothing  in principle would have prevented the number for $\zeta$ extracted from the data from being orders of magnitude off, in which case the present analysis would be invalidated completely.}, and suggests that careful analysis of the data can help to extract non-trivial information about the screening of Glasma flux tubes in p+p collisions. 

\section{Summary and Outlook}

In this work, we extracted impact parameter dependent unintegrated gluon distributions from fits to the HERA inclusive and exclusive data in the IP-Sat, b-CGC and NLO-BK models. These models all implement the physics of saturation but differ in their dynamical assumptions. The impact parameter distributions, in the  $k_\perp$ factorization formalism allow one to compute single inclusive rapidity and $p_{\bot}$ distributions in p+p collisions at the LHC. These give quite reasonable agreement with the LHC data. These impact parameter dependent distributions also allowed us to compute the multiplicity distribution, which in the Color Glass Condensate/Glasma formalism is predicted to be a negative binomial distribution with particular values for the parameter controlling the width of the distribution. We observed that the multiplicity distributions are well reproduced in this framework with numbers for the parameter consistent with values extracted from numerical computations of Yang-Mills equations for the collision of dense color sources. These results suggest that particle emission from Glasma flux tubes generated in collisions of ``hot spots" of  size $1/Q_S$ are a strong candidate for generating the multi-parton correlations underlying the multiplicity distribution. Because these  Glasma flux tubes generate long range rapidity correlations, our results are also consistent with a recent claim~\cite{Dumitru-etal} that Glasma flux tubes generate the near side ridge seen in high multiplicity events by the CMS collaboration~\cite{CMS-ridge}. Because our analysis allows a more careful treatment of the contribution of various impact parameters to the multiplicity distribution, it will allow more quantitative and systematic comparisons to phenomena seen only in high multiplicity cuts in the LHC data. 

There are several caveats in this analysis that must be noted. Firstly, the $k_\perp$ factorization formalism for inclusive multiplicity distributions is rather fragile for $k_\perp \leq Q_S$ because it does not include multi-parton rescatterings~\cite{KrasnV,Lappi}. Significant improvements of the $k_\perp$ factorization formalism are feasible in the CGC framework if presently cumbersome to implement in phenomenological analyses. Similarly, NLO dipole computations are becoming available albeit a consistent NLO dipole analysis still remains to be developed. Both these developments suggest that a combined quantitative study of the collective QCD dynamics of saturation in DIS and hadronic collisions is feasible in future along the lines discussed in this work. 

\section*{Acknowledgements}
R.V was supported by the US Department of Energy under DOE Contract No.DE-AC02-98CH10886. We thank Kevin Dusling and Fran\c{c}ois Gelis for a careful reading of the manuscript and especially Adrian Dumitru for pointing out an error in a previous version of the manuscript. We gratefully acknowledge useful conversations with Javier Albacete, Guillaume Beuf, Subhasis Chattopadhyay, Tuomas Lappi, Larry McLerran, Zhangbo Kang and Feng Yuan.

\end{document}